\documentclass[aps,prd,10pt,twocolumn,preprintnumbers,superscriptaddress,bibnotes,nofootinbib,letterpaper,longbibliography]{revtex4-2}

\usepackage{xcolor}
\usepackage{gensymb}
\definecolor{mygreen}{RGB}{0,108,101}
\definecolor{myorange}{RGB}{255, 128, 1}
\definecolor{myorange}{HTML}{FF7518}
\definecolor{osu}{HTML}{ba0c2f}
\definecolor{ccapp}{HTML}{8a8e3a}
\definecolor{buckeyeblue}{HTML}{001C5A}

\usepackage{graphicx}
\usepackage{amsmath, amssymb}
\usepackage{hyperref}
\usepackage{booktabs}
\usepackage{multirow}
\hypersetup{
  colorlinks  = true,
  urlcolor    = osu,
  linkcolor   = blue,
  citecolor   = red,
}
\usepackage[capitalize]{cleveref}

\newcommand{\orcid}[1]{\begingroup
  \hypersetup{hidelinks}\href{https://orcid.org/#1}{\includegraphics[width=10pt]{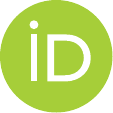}\,} \endgroup}

\begin{document}

\title{Measurement of the neutrino cross section across the TeV scale using IceCube data}

\author{Rasmi Hajjar\,\orcid{0000-0002-9227-5364}}
\email{hajjar.44@osu.edu}
\affiliation{Center for Cosmology and AstroParticle Physics (CCAPP), \href{https://ror.org/00rs6vg23}{\color{osu}The Ohio State University}, Columbus, OH 43210, USA}
\affiliation{Department of Physics, \href{https://ror.org/00rs6vg23}{\color{osu}The Ohio State University}, Columbus, OH 43210, USA}
\affiliation{Department of Astronomy, \href{https://ror.org/00rs6vg23}{\color{osu}The Ohio State University}, Columbus, OH 43210, USA}

\author{Mauricio Bustamante\,\orcid{0000-0001-6923-0865}}
\email{mbustamante@nbi.ku.dk}
\affiliation{Niels Bohr International Academy, Niels Bohr Institute, University of Copenhagen, DK-2100 Copenhagen, Denmark}

\author{Sergio Palomares-Ruiz\,\orcid{0000-0001-9049-2288}}
\email{sergiopr@ific.uv.es}
\affiliation{Instituto de F\'{i}sica Corpuscular (IFIC), CSIC-Universitat de València, E-46980 Paterna, Spain}

\date{\today}

%%%%%%%%%%%%%%%%%%%%%%%%%%%%%%%%%%%%%%%%%%%%%%%%%%%%%%%%%%%%%%%%%%%%%%%%%%%%%%

\begin{abstract}
Neutrinos interact with matter exclusively through the weak force, resulting in extremely small cross sections that are difficult to measure. At TeV energies, neutrino–nucleon interactions probe the structure of the nucleon in regimes where our knowledge of it is less complete, and signatures of physics beyond the Standard Model could manifest. Recent measurements using collider, atmospheric, and astrophysical neutrinos have begun to explore this regime, but no single measurement has provided precise coverage from sub-TeV to tens of TeV, and uncertainties remain at the level of about 20--40\%. Here we present a determination of the muon-neutrino–nucleon charged-current cross sections as a function of energy across $150~\text{GeV}\text{--}40~\text{TeV}$ using eight years of through-going muon neutrino data collected by the IceCube Neutrino Observatory. We measure the average neutrino–antineutrino cross section, enabling direct comparison
with previous results, and find, for the first time above 1~TeV, mild indications that the neutrino and antineutrino cross sections can be measured separately. The measured cross sections are consistent with Standard Model predictions and are determined with uncertainties below 20\% throughout the explored energy range, achieving the most precise characterization of neutrino–nucleon interactions across the TeV scale to date. These results provide a unified measurement of neutrino–nucleon cross sections in this energy regime, connecting accelerator-based and previous higher-energy neutrino observations --- thereby advancing high-energy neutrino physics into the precision era.
\end{abstract}

\maketitle
%%%%%%%%%%%%%%%%%%%%%%%%%%%%%%%%%%%%%%%%%%%%%%%%%%%%%%%%%%%%%%%%%%%%%%%%%%%%%%

%%%%%%%%%%%%%%%%%%%%%%%%%%%%%%%%%%%%%%%%%%%%%%%%
\section{Introduction}
\label{sec:intro}
%%%%%%%%%%%%%%%%%%%%%%%%%%%%%%%%%%%%%%%%%%%%%%%%

The study of nucleon structure through deep inelastic scattering has been a cornerstone of particle physics for over five decades, providing the experimental foundation for our understanding of how quarks and gluons are distributed within protons and neutrons. Neutrino--nucleon scattering, in particular, has played a central role in this effort: the weak force that governs it responds to the various quarks and antiquarks differently from the electromagnetic force that dominates electron and muon scattering, making neutrinos a complementary probe of nucleon structure. At TeV energies and higher, this probe explores regimes where our knowledge of protons and neutrons is less complete~\cite{Conrad:1997ne, Schienbein:2007fs, Muzakka:2022wey, Klasen:2023uqj}, and tests the Standard Model (SM) of particle physics where new phenomena could change how neutrinos interact~\cite{MammenAbraham:2022xoc, Ackermann:2022rqc}.

However, because neutrinos interact exclusively via the weak force, their cross sections with matter are small, so measuring them requires large detectors and high neutrino fluxes. Fixed-target particle accelerator experiments meet both requirements, with intense, well-controlled neutrino beams and massive detectors. They began studying neutrino deep inelastic scattering in the early 1970s using bubble chamber and electronic detectors and, by the late 1970s, had reached neutrino energies of about 200~GeV~\cite{Aachen-Bonn-CERN-London-Oxford-Saclay:1977yve, Aachen-Bonn-CERN-London-Oxford-Saclay:1978eks, Barish:1977ny}. Over the following three decades, statistics and precision improved significantly, but the energy reach grew only modestly, to 360~GeV~\cite{GargamelleSPS:1981hpd, Berge:1987zw, Seligman:1997fe, NuTeV:2005wsg, NOMAD:2007krq}, limited by the energy of the proton beams that produce the neutrinos. This changed dramatically in 2017, when measurements of the cross section with high-energy atmospheric neutrinos (produced by cosmic-ray interactions) and astrophysical neutrinos (of extraterrestrial origin, and first detected in 2013) reached the PeV scale~\cite{IceCube:2017roe, Bustamante:2017xuy, IceCube:2020rnc}. At lower energies, measurements with collider neutrinos first reached $\sim$2~TeV in 2024~\cite{FASER:2024hoe, FASER:2024ref}, and in 2026, Super-Kamiokande measured the cross section at the TeV scale using atmospheric neutrinos~\cite{Super-Kamiokande:2026xsm}.

Despite this recent progress, the energy dependence of the neutrino--nucleon cross section across the TeV scale remains largely unmeasured. The IceCube measurement that spans this range fits a single, energy-independent normalization of the cross section over its entire energy window, 6.3--980~TeV~\cite{IceCube:2017roe}. Super-Kamiokande fits the cross section in two energy bins, but the lower one, below 500~GeV, is constrained by a Gaussian prior from accelerator measurements, so across the TeV scale it measures a single value, averaged over 500~GeV--5~TeV~\cite{Super-Kamiokande:2026xsm}. Energy-resolved measurements of the cross section, on the other hand, stop at $\sim$2~TeV, with collider neutrinos~\cite{FASER:2024hoe, FASER:2024ref}, and resume only at about 10~TeV, with astrophysical neutrinos~\cite{Bustamante:2017xuy, IceCube:2020rnc}. This matters because deviations from the SM prediction, whether from nucleon structure or new physics, would in general depend on energy, and could be diluted in a single normalization fitted over a wide energy range. In addition, uncertainties in measurements above 360~GeV to date remain at the 20--40\% level, several times larger than the uncertainty of the SM prediction~\cite{Cooper-Sarkar:2011jtt}.

Here we measure, as a function of energy from 150~GeV to 40~TeV, the neutrino cross section for charged-current interactions, in which a muon neutrino or antineutrino scatters off a nucleon and turns into a muon. We use a single dataset: eight years of public IceCube Neutrino Observatory data on upward-going neutrinos that reach the detector from below, after traveling through the Earth~\cite{IceCube:2020phf, IceCube:2020tka, IceCube:2023gdv, DVN/9WGYQN_2024}. Our measurement improves on previous work in three ways. First, our  binning in energy spans the gap between the collider and astrophysical results, and overlaps with both. Second, despite this finer binning, the uncertainty on the cross section in every bin is below 20\%, comparable to or smaller than that of previous measurements above 360~GeV. Third, we find a first, albeit mild, indication that the neutrino and antineutrino cross sections can be measured separately above 1~TeV, even though IceCube cannot tell neutrinos from antineutrinos event by event.

%%%%%%%%%%%%%%%%%%%%%%%%%%%%%%%%%%%%%%%%%%%%%%%%
\section{The IceCube data sample}
\label{sec:ICdata}
%%%%%%%%%%%%%%%%%%%%%%%%%%%%%%%%%%%%%%%%%%%%%%%%

The IceCube Neutrino Observatory instruments a cubic kilometer of ice at the South Pole with 5,160 light sensors, called digital optical modules, at depths between 1.45 and 2.45~km~\cite{IceCube:2016zyt}. The dataset we use was collected during 2011--2019~\cite{IceCube:2020phf, IceCube:2020tka, IceCube:2023gdv, DVN/9WGYQN_2024} and contains 305,735 selected events reconstructed in muon energy and zenith angle, $\theta_z$, measured from the position of the detector. These events consist of tracks of light produced when upward-going muon neutrinos and antineutrinos of atmospheric or astrophysical origin that traverse the Earth undergo charged-current interactions inside the detector or in the surrounding bedrock or ice. The resulting muons travel through the detector at speeds exceeding the speed of light in ice, emitting Cherenkov radiation that is detected by the digital optical modules. At TeV energies, muons can cross the entire detector before falling below the energy threshold for Cherenkov emission, leaving a characteristic extended track signature.

We use the associated IceCube data release that also includes the signal simulation and  details about the treatment of systematic uncertainties~\cite{IceCube:2023gdv, DVN/9WGYQN_2024}. The event selection has a purity higher than 99.9\% in muon neutrinos and antineutrinos, and 99\% of all these events originated from atmospheric or astrophysical neutrinos and antineutrinos with energies between 150~GeV and 40~TeV. Further information on the data sample and the event rate modeling is provided in Appendix~\ref{ap:ICsample}.

%%%%%%%%%%%%%%%%%%%%%%%%%%%%%%%%%%%%%%%%%%%%%%%%
\section{Neutrino--nucleon interactions at TeV energies}
\label{sec:TeVint}
%%%%%%%%%%%%%%%%%%%%%%%%%%%%%%%%%%%%%%%%%%%%%%%%

Neutrinos scatter off nucleons through two channels. In charged-current interactions, mediated by the exchange of a $W$ boson, the neutrino converts into a charged lepton of the same flavor (in the IceCube sample, a muon or an antimuon) that, at TeV energies, carries away most of the available energy. In neutral-current interactions, mediated by the exchange of a $Z$ boson, the neutrino remains a neutrino and transfers only part of its energy to the target. These interactions, whose cross section at these energies is about a third of the charged-current one, leave no track and do not enter the event selection, but they are relevant for the flux that reaches the detector, since they degrade the energy of the neutrinos as they propagate through the Earth.

Above a few tens of GeV, both channels proceed via deep inelastic scattering: the exchanged boson resolves the individual partons (quarks and gluons) inside the nucleon, which breaks up into a hadronic shower. The cross section then follows from convolving the elementary neutrino-parton cross section with the parton distribution functions, evaluated at the fraction of the nucleon momentum carried by the struck parton --- the Bjorken-$x$ variable --- and at the negative of the squared momentum transfer, $Q^2$~\cite{Conrad:1997ne}. In the IceCube sample, however, neutrinos scatter off nucleons bound in the nuclei of the Earth and of the ice. Two nuclear effects can make the cross section on them differ from the one on free nucleons, but both are small compared to our uncertainties. 

First, nuclei generally contain unequal numbers of protons and neutrons, whereas calculations usually assume an isoscalar target, with equal numbers of both. This imbalance matters when neutrinos scatter mainly off valence quarks, which differ between protons (quark content $uud$) and neutrons ($udd$): in charged-current interactions, neutrinos scatter off down quarks ($d$) and antineutrinos off up quarks ($u$). As the neutrino energy increases, however, the cross section is increasingly dominated by small-$x$ scattering off sea quarks and antiquarks, which are nearly the same in protons and neutrons. We have explicitly checked that the isoscalar approximation implies an effect of just a few percent and, thus, it has negligible impact on our results (see Appendix~\ref{ap:robustness}). Second, the structure of bound nucleons differs from that of free nucleons~\cite{Schienbein:2007fs, Muzakka:2022wey, Klasen:2023uqj}. Nevertheless, at the neutrino energies relevant here ($E_\nu < 40~\text{TeV}$), these nuclear corrections are expected to increase the cross section by 1--2\%~\cite{Klein:2020nuk}, well below the current uncertainties, too. Therefore, we treat the Earth and the detector as isoscalar targets and report our cross sections, like the CSMS ones that we use as reference~\cite{Cooper-Sarkar:2011jtt}, per free nucleon.

In addition to $E_\nu$, $x$, and $Q^2$, another important kinematic variable of the interaction is the inelasticity, $y$, the fraction of the neutrino energy transferred to the hadronic system; it is related to the other variables by $y = Q^2/\left(2 \, M \, E_\nu \, x\right)$, where $M$ is the nucleon mass. The outgoing muon carries the remaining energy, $(1-y) \, E_\nu$. This makes the inelasticity particularly important for our analysis: IceCube reconstructs the muon energy, not the neutrino energy, and the two are connected by the broad distribution in $y$ and, for muons produced outside the detector, by the energy they lose before reaching it. The mean inelasticity is about 0.5 for neutrinos and 0.35 for antineutrinos at a few hundred GeV, and decreases to about 0.4 and 0.3, respectively, at 40~TeV. The inelasticity distribution also underlies the difference between the neutrino and antineutrino cross sections. At accelerator energies, where the scattering is dominated by valence up and down quarks, the distribution in $y$ is nearly flat for neutrinos and strongly suppressed towards large $y$ for antineutrinos, so that the antineutrino cross section is roughly half of the neutrino one. As the energy grows, the contributions from sea quarks and antiquarks take over and the two cross sections approach each other, a convergence already well under way at the highest energies in the IceCube sample. Measurements with atmospheric and astrophysical neutrinos have found the inelasticity distribution to be consistent with standard calculations~\cite{IceCube:2018pgc, IceCubeCollaborationSS:2025zgz}; as in those works, we adopt the CSMS inelasticity distribution~\cite{Cooper-Sarkar:2011jtt}.

Nuclear effects could also modify the inelasticity distribution, increasing the fraction of low-inelasticity interactions, particularly below a few TeV~\cite{Klein:2020nuk, SajjadAthar:2020nvy, Candido:2023utz}. We neglect them here, for two reasons. First, the inelasticity can only be measured directly when both the hadronic shower and the muon are seen, which requires the neutrino to interact inside the detector; most events in the IceCube sample interact outside it, so only the muon is observed. Second, nuclear effects mainly shift interactions towards low inelasticity, where the muon already takes most of the neutrino energy. Since the muon energy is $(1-y) \, E_\nu$, a change $\Delta y$ in the inelasticity shifts it by $\Delta y \, E_\nu$, which is small when $\Delta y$ is small, even if the relative change, $\Delta y / y$, is large. This shift is smaller than the uncertainty with which the muon energy is reconstructed (see Appendix~\ref{ap:ICsample}).

\vspace{-1.08ex}
%%%%%%%%%%%%%%%%%%%%%%%%%%%%%%%%%%%%%%%%%%%%%%%%
\section{Sensitivity to the neutrino--nucleon cross section}
\label{sec:sensitivity}
%%%%%%%%%%%%%%%%%%%%%%%%%%%%%%%%%%%%%%%%%%%%%%%%

We measure the cross section through its effect on the distribution of detected events in reconstructed muon energy and zenith angle, which it shapes in two ways. First, for given incident neutrino and antineutrino fluxes at the detector, $\Phi_\nu$ and $\Phi_{\bar{\nu}}$, the rate of detected events is proportional to the combination of the neutrino and antineutrino cross sections, $N_{\rm events} \propto \sigma_{\nu} \, \Phi_\nu + \sigma_{\bar\nu} \, \Phi_{\bar{\nu}}$. Second, the cross section determines how much of the upward-going flux the Earth absorbs before it reaches the detector, which distorts the energy and zenith-angle distributions of the events. We exploit both effects jointly, fitting the cross section in each energy bin together with the parameters that describe the fluxes and the detector response (see Appendix~\ref{ap:parameter_est}).

Absorption in the Earth becomes important at high energies because the probability of neutrino interaction with matter grows with energy. In a simplified picture, muon neutrinos of energy $E_\nu$ that travel a distance $L(\theta_z)$ through the Earth along a chord with zenith angle $\theta_z$ reach the detector with a flux $\Phi_{\nu_\mu}(E_\nu, \theta_z) \simeq \Phi^0_{\nu_\mu}(E_\nu, \theta_z) \, e^{-L(\theta_z)/\lambda_{\nu_\mu}(E_\nu, \theta_z)}$, where $\Phi^0_{\nu_\mu}$ is the flux before entering the Earth, $\lambda_{\nu_\mu} = [(\sigma^{\rm CC}_{\nu_\mu} + \sigma^{\rm NC}_{\nu_\mu}) \, n_N]^{-1}$ is the interaction length, and $n_N(\theta_z)$ is the average number density of nucleons along the path; for muon antineutrinos, the antineutrino cross sections enter instead. For 40-TeV neutrinos, the interaction length is of the order of the Earth's diameter, so above a few TeV the Earth is not fully transparent to neutrino propagation, and the attenuation grows with both the neutrino energy and the distance traveled, which is set by the zenith angle. This exponential form is only illustrative: our calculation solves the full transport equations, which also include the regeneration of neutrinos at lower energies by neutral-current interactions, as well as flavor oscillations (see Appendix~\ref{ap:Earth_prop}). The IceCube sample spans a wide range of baselines through the Earth (from a few to thousands of kilometers) and energies (from 150~GeV to 40~TeV), covering both energies at which the Earth is transparent and energies at which it absorbs part of the flux.

Because the event rate depends on the product of flux and cross section, the two are, in principle, degenerate and cannot be determined independently from each other. Absorption breaks this degeneracy, because it changes the contribution from neutrinos that cross different amounts of Earth~\cite{Hooper:2002yq}. Furthermore, the energy distribution of the events provides a crucial handle to disentangle the cross sections from the flux, because the flux varies smoothly with energy, whereas we let the cross section vary independently in each energy bin. In addition, the flux of atmospheric neutrinos, which dominates the sample, is independently constrained by measurements of the cosmic-ray spectrum, by the hadronic interaction models describing the air showers in which these neutrinos originate, and by the correlation with the atmospheric muon flux produced from the same decay chain of charged mesons. The uncertainty on the muon neutrino and antineutrino fluxes stays below 10\% up to 1~TeV and increases to 25\% at the highest energies relevant to this analysis, whereas the uncertainties on the muon neutrino--antineutrino flux ratio are below 8\% in the entire energy range~\cite{Yanez:2023lsy}.

%%%%%%%%%%%%%%%%%%%%%%%%%%%%%%%%%%%%%%%%
\begin{figure*}[t]
    \begin{center}
  		\includegraphics[width=\textwidth]{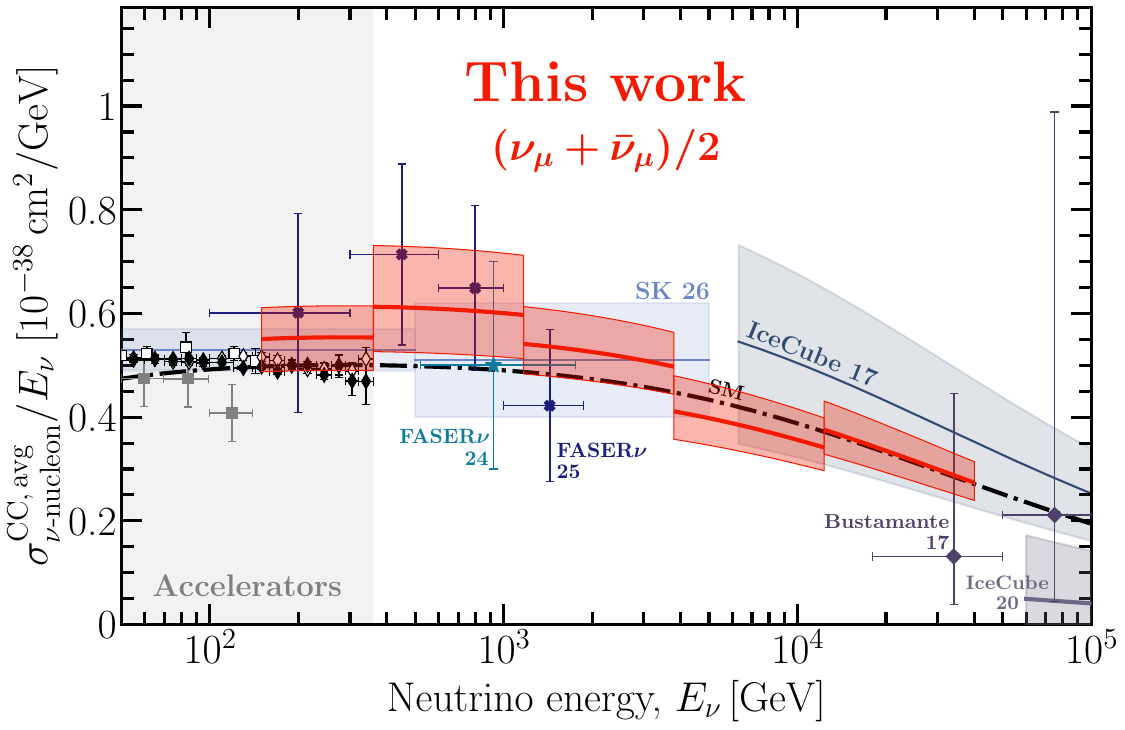}
  	    \caption{
            \textbf{Measurements of the average muon neutrino--antineutrino charged-current cross section}, fixing their ratio to the SM prediction. Solid red lines show the modes, and shaded red bands the 68.27\% highest probability density (HPD) intervals, for five energy bins in the range $E_\nu \in [150~\text{GeV}, \, 40~\text{TeV}]$, using the IceCube eight-year dataset of upward-going muon neutrinos~\cite{IceCube:2020phf, IceCube:2020tka, IceCube:2023gdv, DVN/9WGYQN_2024}. We include a Gaussian prior with a 16\% relative uncertainty for the first bin, $E_1 \in [150, \, 360]~\text{GeV}$. 
            We also show results from fixed-target~\cite{GargamelleSPS:1981hpd, Berge:1987zw, Seligman:1997fe, NuTeV:2005wsg}, collider~\cite{FASER:2024hoe, FASER:2024ref}, and atmospheric and astrophysical~\cite{IceCube:2017roe, Bustamante:2017xuy, IceCube:2020rnc, Super-Kamiokande:2026xsm} neutrinos (see the text). The Super-Kamiokande results (``SK 26'') are from Table~I in~\cite{Super-Kamiokande:2026xsm}. We show the IceCube 17 result, $1.30^{+0.21}_{-0.19}~\text{(stat.)}~^{+0.39}_{-0.43}~\text{(syst.)}$ times the CSMS prediction of the SM~\cite{Cooper-Sarkar:2011jtt}, which is represented by the dash-dotted line.
  	   }	
    \label{fig:xsec_mean_result}
    \end{center}
\end{figure*}
%%%%%%%%%%%%%%%%%%%%%%%%%%%%%%%%%%%%%%%%

The atmospheric neutrino flux is split into a conventional component, produced by the decay of pions and kaons with a neutrino energy spectrum that falls as $\sim E_\nu^{-3.7}$, and a subdominant prompt component from charmed hadron decays with an energy spectrum that falls as $\sim E_\nu^{-2.7}$. The latter has not been experimentally identified yet and its normalization is unknown. Nevertheless, its contribution to the IceCube sample is expected to be small, at the 0.1\% level~\cite{IceCube:2020tka}, so we do not include it in the analysis. Additionally, astrophysical neutrinos, produced in energetic sources beyond the Solar System, could represent a significant contribution at the highest energies in this analysis, depending on their spectrum, which we model as an unbroken single power law, $\propto E_\nu^{-\gamma_{\rm astro}}$, with free normalization, $\Phi_{\rm astro}$, and spectral index, $\gamma_{\rm astro}$ (see Appendix~\ref{ap:nu_flux}).

%%%%%%%%%%%%%%%%%%%%%%%%%%%%%%%%%%%%%%%%
\begin{table}[t] 
	\centering
		\begin{tabular*}{\columnwidth}{@{\extracolsep{\fill}}ccc}
		\hline
		Energy range [TeV] & Prior & Normalization \\
		\hline \\
		0.15 -- 0.36  & $\mathcal{N}(1.00,0.16)$    & $1.11^{+0.12}_{-0.13}$     \\[2ex]
		0.36 -- 1.17    & $\mathcal{U}(0,20)$     & $1.22^{+0.24}_{-0.17}$     \\[2ex]
		1.17 -- 3.79  & $\mathcal{U}(0,20)$     & $1.11^{+0.15}_{-0.12}$     \\[2ex]
		3.79 -- 12.32    & $\mathcal{U}(0,20)$     & $0.92^{+0.15}_{-0.12}$     \\[2ex]
		12.32 -- 40    & $\mathcal{U}(0,20)$     & $1.01^{+0.15}_{-0.13}$     \\[2ex]
		\hline
	\end{tabular*}
	\caption{ \textbf{Average muon neutrino--antineutrino cross section, fixing their ratio to the SM prediction.} Modes of the posteriors and 68.27\% HPD intervals for the five neutrino energy bins are indicated. All normalizations are obtained with respect to the CSMS average muon neutrino--antineutrino cross section in each energy bin~\cite{Cooper-Sarkar:2011jtt}. $\mathcal{N}(m, s)$ denotes a Gaussian prior with mean $m$ and standard deviation $s$, and $\mathcal{U}(a, b)$ a uniform prior between $a$ and $b$. (The cross sections we report in this table and everywhere else in the paper are not weighted by the neutrino flux.)}
	\label{tab:normalization_1_5}
\end{table}
%%%%%%%%%%%%%%%%%%%%%%%%%%%%%%%%%%%%%%%%

%%%%%%%%%%%%%%%%%%%%%%%%%%%%%%%%%%%%%%%%%%%%%%%%
\section{Average neutrino and antineutrino cross sections}
\label{sec:result_average}
%%%%%%%%%%%%%%%%%%%%%%%%%%%%%%%%%%%%%%%%%%%%%%%%

As in previous measurements with atmospheric and astrophysical neutrinos~\cite{IceCube:2017roe, Bustamante:2017xuy, IceCube:2020rnc, Super-Kamiokande:2026xsm}, we vary only the normalization of the neutrino and antineutrino cross sections in each energy bin, keeping the shape of the differential cross sections, and hence the inelasticity distribution, unchanged. The same normalizations are applied to both charged-current and neutral-current cross sections, and we treat the Earth and the detector as isoscalar targets.

To enable direct comparison with previous results, we first scale the neutrino and antineutrino cross sections by the same factor in each energy bin (i.e., we fix their ratio to the SM prediction), and report their average, $(\sigma_\nu^{\rm CC} + \sigma_{\bar\nu}^{\rm CC})/2$. We characterize this cross section by splitting the neutrino energy range, $E_\nu \in [150~\text{GeV}, \, 40~\text{TeV}]$, into five bins. The first bin comprises neutrino energies where fixed-target measurements exist, $E_1 \in [150, \, 360]~\text{GeV}$. The other four bins are chosen to have the same logarithmic widths. We allow the normalization of each bin to vary independently with respect to the CSMS cross section~\cite{Cooper-Sarkar:2011jtt}, which we adopt as the reference. We use a Gaussian prior only for the first bin. Although the uncertainties from measurements at fixed-target experiments are of a few percent~\cite{GargamelleSPS:1981hpd, Berge:1987zw, Seligman:1997fe, NuTeV:2005wsg, NOMAD:2007krq}, we conservatively set the standard deviation of this prior to 16\%, following the same prescription as Super-Kamiokande~\cite{Super-Kamiokande:2026xsm}. This approximately corresponds to the total uncertainty on the last energy bin of the NuTeV analysis~\cite{NuTeV:2005wsg}. Together with the atmospheric flux model, this prior helps constrain the flux normalization, which is common to all energies, and, through it, the cross section in the other bins. Without it, the fit prefers a lower flux normalization, and the cross-section normalizations in the other four bins shift upward by 20--30\% (see Appendix~\ref{ap:robustness}).

We include the same set of systematic uncertainties as in the original IceCube analyses that introduced the data sample we use~\cite{IceCube:2020phf, IceCube:2020tka, IceCube:2023gdv}, which account for uncertainties in the primary cosmic-ray flux, the hadronic interaction models governing air-shower development, the astrophysical neutrino flux, the detector response, and the optical properties of the ice (see Appendix~\ref{ap:systematics}). These systematic uncertainties enter the analysis as nuisance parameters in a Bayesian fit based on a Poisson likelihood, where they are continuously varied and marginalized simultaneously with the parameters of interest (i.e., the cross-section normalizations in each energy bin). 

The results of this analysis are shown in Fig.~\ref{fig:xsec_mean_result}, along with previous measurements from fixed-target, collider, atmospheric, and astrophysical neutrinos. Only measurements using atmospheric and astrophysical neutrinos~\cite{IceCube:2017roe, Bustamante:2017xuy, IceCube:2020rnc, Super-Kamiokande:2026xsm}, and the highest-energy results from FASER~\cite{FASER:2024hoe, FASER:2024ref}, are reported as the average muon neutrino--antineutrino cross section. For fixed-target~\cite{GargamelleSPS:1981hpd, Berge:1987zw, Seligman:1997fe, NuTeV:2005wsg} and collider~\cite{FASER:2024ref} neutrinos below 1~TeV, which report separate neutrino and antineutrino cross sections, we perform the average ourselves, adding their separate errors in quadrature. 

Our results represent the most precise measurement of the neutrino--nucleon cross section across the TeV scale to date, covering the range from sub-TeV to tens of TeV energies. The relative uncertainty in every energy bin is smaller than 20\% of the measured value and the results are consistent with the SM expectation. Table~\ref{tab:normalization_1_5} lists the cross-section normalizations.

%%%%%%%%%%%%%%%%%%%%%%%%%%%%%%%%%%%%%%%%
\begin{figure*}[t]
    \begin{center}
        \includegraphics[width=\textwidth]{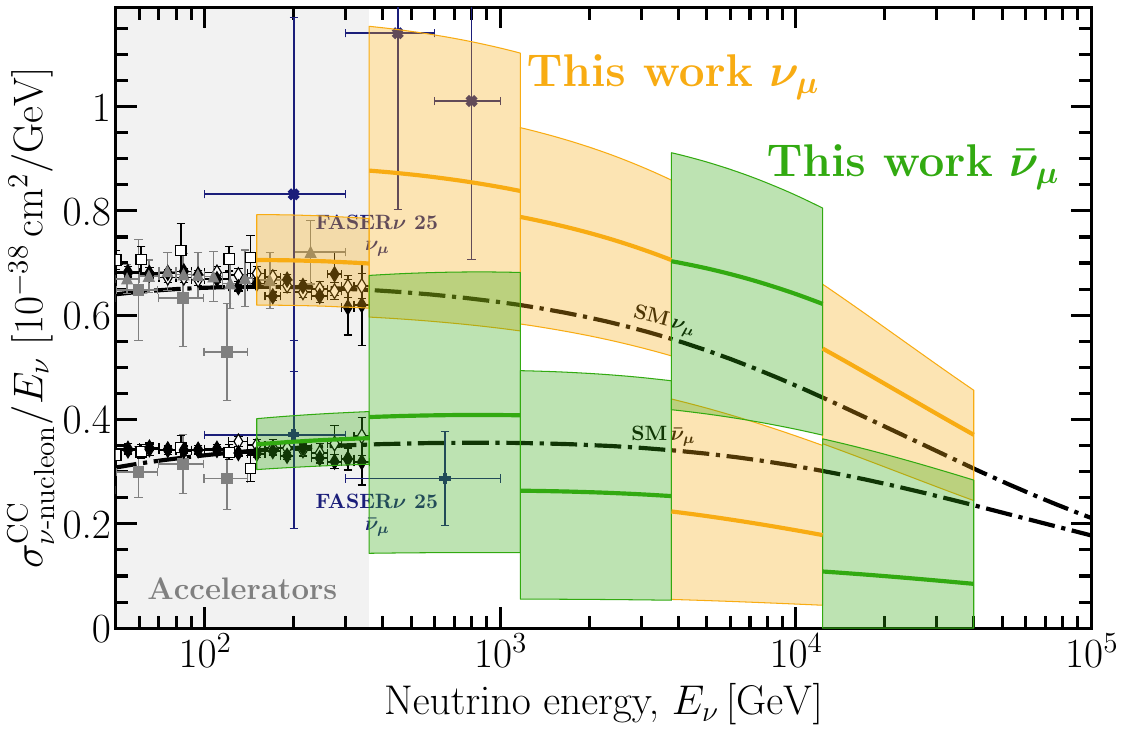}
        \caption{\textbf{Measurement of the muon neutrino and antineutrino charged-current cross sections.}  In orange (neutrinos) and green (antineutrinos), the mode and the 68.27\% HPD interval for five energy bins in the range $E_\nu \in [150~\text{GeV}, \, 40~\text{TeV}]$, using the IceCube eight-year dataset of upward-going muon neutrinos~\cite{IceCube:2020phf, IceCube:2020tka, IceCube:2023gdv, DVN/9WGYQN_2024}. For both cross sections, we include a Gaussian prior with a 16\% relative uncertainty for the first bin, $E_1 \in [150, \, 360]~\text{GeV}$. We also show results from fixed-target~\cite{GargamelleSPS:1981hpd, Berge:1987zw, Seligman:1997fe, NuTeV:2005wsg, NOMAD:2007krq} and collider~\cite{FASER:2024ref} neutrinos for $E_\nu < 1~\text{TeV}$. This is the first mild indication of a separate measurement of neutrino and antineutrino cross sections above 1~TeV; the significance of a non-zero antineutrino cross section for $E_\nu > 360~\text{GeV}$ is $1.6\sigma$. The dash-dotted lines represent the theoretical estimates for the CSMS neutrino and antineutrino cross sections~\cite{Cooper-Sarkar:2011jtt}.}
    \label{fig:xsec_nunb_result}
    \end{center}
\end{figure*}
%%%%%%%%%%%%%%%%%%%%%%%%%%%%%%%%%%%%%%%%

%%%%%%%%%%%%%%%%%%%%%%%%%%%%%%%%%%%%%%%%
\begin{table}[t]  
	\centering
        \begin{tabular*}{\columnwidth}{@{\extracolsep{\fill}}cccc}
		\hline
		Energy range [TeV] & Prior & \multicolumn{2}{c}{Normalization} \\
		\hline \\
		\multirow{2}{*}{0.15 -- 0.36}  & \multirow{2}{*}{$\mathcal{N}(1.00,0.16)$}       & $\nu:$ & $1.08 \pm 0.13$  \\
		&  & $\bar\nu:$ & $1.03 \pm 0.14$     \\[2ex]
		\multirow{2}{*}{0.36 -- 1.17}    & \multirow{2}{*}{$\mathcal{U}(0,20)$}     & $\nu:$ & $1.4 \pm 0.4$  \\
		& & $\bar\nu:$ & $1.2^{+0.8}_{-0.7}$     \\[2ex]
        \multirow{2}{*}{1.17 -- 3.79}    & \multirow{2}{*}{$\mathcal{U}(0,20)$}     & $\nu:$ & $1.27^{+0.28}_{-0.33}$  \\
		& & $\bar\nu:$ & $0.7^{+0.7}_{-0.6}$     \\[2ex]
        \multirow{2}{*}{3.79 -- 12.32}    & \multirow{2}{*}{$\mathcal{U}(0,20)$}     & $\nu:$ & $0.40^{+0.39}_{-0.30}$  \\
		& & $\bar\nu:$ & $2.1^{+0.6}_{-0.8}$     \\[2ex]
        \multirow{2}{*}{12.32 -- 40}    & \multirow{2}{*}{$\mathcal{U}(0,20)$}     & $\nu:$ & $1.21^{+0.28}_{-0.41}$  \\
		& & $\bar\nu:$ & $0.4^{+0.8}_{-0.4 \, \, \dagger}$     \\[2ex]
		\hline
	\end{tabular*}
	\caption{\textbf{Separate muon neutrino and antineutrino cross sections.} Modes of the posteriors and 68.27\% HPD intervals for the five neutrino energy bins are indicated. All normalizations are obtained with respect to the CSMS muon neutrino and antineutrino cross sections in each energy bin~\cite{Cooper-Sarkar:2011jtt}. Priors are denoted as in Table~\ref{tab:normalization_1_5}. $^\dagger$ The interval reaches the physical bound at 0.}
	\label{tab:nunub_normalization_1_5}
\end{table}
%%%%%%%%%%%%%%%%%%%%%%%%%%%%%%%%%%%%%%%%

%%%%%%%%%%%%%%%%%%%%%%%%%%%%%%%%%%%%%%%%%%%%%%%%
\section{Separate neutrino and antineutrino cross sections}
\label{sec:result_nu_and_nubar}
%%%%%%%%%%%%%%%%%%%%%%%%%%%%%%%%%%%%%%%%%%%%%%%%

The IceCube detector does not identify the charge of the muon, so neutrinos and antineutrinos cannot be told apart on an event-by-event basis. Their cross sections could, in principle, be disentangled statistically, because the two species differ in three ways that shape the observed distributions: in their fluxes, in their inelasticity distributions, and in their absorption in the Earth. First, the atmospheric flux contains fewer antineutrinos than neutrinos, in a ratio that is predicted with an uncertainty below 8\% over the energy range of this analysis~\cite{Yanez:2023lsy}. Second, the inelasticity distribution of antineutrinos is weighted towards smaller values of $y$, so that, for the same neutrino energy, the outgoing antimuon carries a larger fraction of the energy than the muon produced by a neutrino, and the reconstructed muon energy distributions of the two species are displaced relative to each other. Third, because the antineutrino cross section is smaller, antineutrinos are less attenuated while traversing the Earth, and the two species therefore have different energy and zenith-angle distributions, which thus retain information on each species separately.

To statistically disentangle the neutrino and antineutrino contributions, we consider the same five energy bins used in the previous analysis, but now allow the normalizations of the neutrino and antineutrino cross sections within these bins to vary independently of each other. As above, the normalizations are defined with respect to the CSMS calculations in each bin~\cite{Cooper-Sarkar:2011jtt}. We also apply a 16\% Gaussian prior to the normalization of the first bin ($E_1 \in [150, 360]~\text{GeV}$) for each species, while adopting uniform priors elsewhere. All other components of the fit, including the treatment of nuisance parameters, remain unchanged.
 
The results are presented in Fig.~\ref{fig:xsec_nunb_result} and summarized in Table~\ref{tab:nunub_normalization_1_5}, alongside fixed-target and collider measurements available below 1~TeV. To our knowledge, no separate neutrino and antineutrino cross sections have been reported at higher energies. In the first bin, both cross sections are measured in agreement with their SM values, with uncertainties of 12\% for neutrinos and 14\% for antineutrinos, comparable to the precision of our average cross section measurement in the same bin. This is primarily driven by the Gaussian prior imposed from fixed-target results, which is crucial for constraining the overall atmospheric neutrino and antineutrino flux normalization. For $E_\nu > 360~\text{GeV}$, the measured neutrino cross-section normalizations are consistent with SM expectations within $1\sigma$, except for the fourth bin, which exhibits an underfluctuation slightly exceeding $1\sigma$. The antineutrino cross-section normalizations are considerably less precise, as expected from the smaller antineutrino event fraction, yet they also align with SM expectations, save for an overfluctuation in the fourth bin compensating for the neutrino deficit.
   
Current data do not yet allow a definitive measurement of the antineutrino cross section above $360~\text{GeV}$. Comparing the marginalized posteriors of this analysis with those obtained by setting the antineutrino normalizations in the last four energy bins to zero reveals a slight preference for a non-zero antineutrino cross section for $E_\nu > 360~\text{GeV}$ with a $p$-value of 0.06. This corresponds to $1.6\sigma$ (one-sided) significance, assuming the log-likelihood ratio behaves as a weighted mixture of $\chi^2$ distributions with up to four degrees of freedom~\cite{Chernoff1954}. We have also evaluated how much of this significance is driven by the overfluctuation in the fourth bin by forcing the neutrino cross section to be larger than the antineutrino one. In this case, the peaks in the posterior distributions for that bin are swapped, making the antineutrino cross section smaller than the neutrino one, and reducing the significance for a non-zero antineutrino cross section to $1.1\sigma$ ($p$-value = 0.14).

We have also estimated the separation of the neutrino and antineutrino cross sections using the Mahalanobis distance of the four highest-energy bins, which results in a mild $1.3\sigma$ separation. Although mild, this separation is  the first step towards a separate determination of the muon neutrino and antineutrino cross sections using atmospheric or astrophysical neutrinos, and the first of any kind reported above 1~TeV. Below 1~TeV they agree with the fixed-target~\cite{GargamelleSPS:1981hpd, Berge:1987zw, Seligman:1997fe, NuTeV:2005wsg, NOMAD:2007krq} and collider~\cite{FASER:2024ref} results obtained with beams of known composition, although in the first bin this agreement is aided by the prior. Above 1~TeV, where the two species have never been resolved, they provide a mild indication that a neutrino telescope can measure them separately, even without charge identification.

%%%%%%%%%%%%%%%%%%%%%%%%%%%%%%%%%%%%%%%%%%%%%%%%
\section{Summary and concluding remarks}
\label{sec:summary}
%%%%%%%%%%%%%%%%%%%%%%%%%%%%%%%%%%%%%%%%%%%%%%%%

We have measured the muon neutrino charged-current cross section from 150~GeV to 40~TeV as a function of energy, using a single dataset, the public eight-year sample of upward-going muon neutrinos from IceCube~\cite{IceCube:2020phf, IceCube:2020tka, IceCube:2023gdv, DVN/9WGYQN_2024}. Previous measurements across this range each obtained a single value of the cross section over the TeV scale, while energy-resolved measurements existed only below $\sim$2~TeV and above about 10~TeV. Our work bridges the two, overlapping with both, with relative uncertainties below 20\% in every energy bin, and agrees with the SM prediction throughout. We have also found a first, albeit mild, indication that the neutrino and antineutrino cross sections can be measured separately above 1~TeV. Although current data do not yet allow a definitive measurement of a non-zero antineutrino cross section above 360~GeV, a neutrino telescope could in principle tell the two apart, even without identifying the charge of the muon. 

These results have two consequences. First, analyses of high-energy neutrinos, whether they target the atmospheric and astrophysical fluxes, their flavor composition, or signatures of new physics, depend on the neutrino--nucleon cross section, which they have so far assumed to be given by theoretical calculations. Our work tests this input directly across the TeV scale. Second, the agreement with the SM supports the parton distribution functions on which these calculations rest, and leaves room only for deviations, such as those predicted in some scenarios of physics beyond the SM, no larger than the present uncertainties. All of this was achieved using only publicly released IceCube data~\cite{IceCube:2020phf, IceCube:2020tka, IceCube:2023gdv, DVN/9WGYQN_2024} and conservative constraints from previous measurements below 360~GeV~\cite{GargamelleSPS:1981hpd, Berge:1987zw, Seligman:1997fe, NuTeV:2005wsg, NOMAD:2007krq}.

The precision of the measurement is limited mainly by the degeneracy between the cross section and the normalization of the atmospheric neutrino flux, which the present sample breaks only partially. Larger samples, already being accumulated, could constrain the flux better and thereby reduce this degeneracy, improving the precision of the cross section; they could also sharpen the separate measurement of the neutrino and antineutrino cross sections. With ten times the present exposure, within reach of next-generation neutrino telescopes, the uncertainty would shrink below the 10\% level on the average cross section, assuming the neutrino--antineutrino cross section ratio to be the SM one. For separate measurements of the neutrino and antineutrino cross sections, uncertainties would shrink below the 20\% level for neutrinos and below the 40\% level for antineutrinos. Remarkably, a non-zero antineutrino cross section above 360~GeV could be established at more than $7\sigma$ (see Appendix~\ref{ap:forecast}). Between 360~GeV and $\sim$2~TeV, such measurements can be compared directly with those made with collider neutrinos~\cite{FASER:2024hoe, FASER:2024ref}, testing the same quantity with two independent neutrino sources. Already with current IceCube data, the uncertainties in this energy interval are smaller than those obtained with collider neutrinos.

The same strategy, inferring the cross section from the absorption of neutrinos inside the Earth, extends to ultra-high energies, above 100~PeV. There, the Earth is opaque to neutrinos except along nearly horizontal directions, and upcoming neutrino telescopes could measure the cross section at energies more than a thousand times higher than those reached here~\cite{Denton:2020jft, Valera:2022ylt, Esteban:2022uuw}, probing the structure of the nucleon beyond the reach of accelerator experiments. Unlike at TeV energies, where the atmospheric neutrino flux is constrained independently, the ultra-high-energy neutrino flux is unknown and must be measured together with the cross section~\cite{Valera:2023ayh}. 

The present work demonstrates, with data from a neutrino telescope operating today, the approach on which those measurements rely: extracting the cross section from the energy and angular distributions of Earth-crossing neutrinos while marginalizing over the uncertainties of their flux.

%%%%%%%%%%%%%%%%%%%%%%%%%%%%%%%%%%%%%%%%%%%%%%%%%%%%%%%%%%%%%%%%%%%%%%%%%%%%%%
\begin{acknowledgments}
The authors thank Alex Wen and Carlos Argüelles for help with the IceCube data release, and Jordi Salvadó for help with the \texttt{nuSQuIDS} code. RH would like to thank the Niels Bohr Institute for hospitality during the completion of this work. MB is supported by {\sc Villum Fonden} under project no.~29388. SPR is supported by grant PID2023-151418NB-I00, funded by MICIU/AEI/10.13039/501100011033/ and by FEDER, UE; by SO project CEX2023-001292-S, funded by MICIU/AEI; by the Generalitat Valenciana grant CIPROM/2022/36; and by the European Union’s Horizon 2020 research and innovation programme under the Marie Skłodowska-Curie grant HORIZON-MSCA-2021-SE-01/101086085-ASYMMETRY.
\end{acknowledgments}

%%%%%%%%%%%%%%%%%%%%%%%%%%%%%%%%%%%%%%%%%%%%%%%%%%%%%%%%%%%%%%%%%%%%%%%%%%%%%%
%%%%%%%%%%%%%%%%%%%%%%%%%%%%%%%%%%%%%%%%%%%%%%%%%%%%%%%%%%%%%%%%%%%%%%%%%%%%%%

\clearpage
\onecolumngrid
\appendix
\renewcommand\thefigure{\thesection\arabic{figure}}
\renewcommand\thetable{\thesection\arabic{table}}
\setcounter{figure}{0}
\setcounter{table}{0}

{\centering\huge\textsc{Appendices}\par}
\vspace{2em}

Below, we describe the key aspects of our statistical analysis. Appendix~\ref{ap:ICsample} details the dataset used in this work. Appendix~\ref{ap:nu_flux} describes the atmospheric and astrophysical neutrino fluxes. Appendix~\ref{ap:Earth_prop} outlines our treatment of neutrino propagation through the Earth. Details of the statistical framework are provided in Appendix~\ref{ap:parameter_est}, while the impact of systematic uncertainties is examined in Appendix~\ref{ap:systematics}. In Appendix~\ref{ap:robustness}, we present additional tests validating our approach and conclusions. Finally, Appendix~\ref{ap:forecast} estimates future sensitivity to the neutrino--nucleon cross section across the TeV scale using mock data.

%%%%%%%%%%%%%%%%%%%%%%%%%%%%%%%%%%%%%%%%%%%%%%%%
\section{IceCube eight-year data sample of upward-going muon neutrinos}
\label{ap:ICsample}
%%%%%%%%%%%%%%%%%%%%%%%%%%%%%%%%%%%%%%%%%%%%%%%%

The eight-year neutrino data sample exploited in this work was initially used for updating the high-energy sterile neutrino searches performed by the IceCube Collaboration~\cite{IceCube:2020phf, IceCube:2020tka}. It was publicly released~\cite{DVN/9WGYQN_2024} accompanying a study of neutrino decoherence~\cite{IceCube:2023gdv}. This sample represents an improved event selection over the earlier dataset~\cite{IceCube:2016rnb}, not only in statistics, but also in the treatment of systematic uncertainties and statistical methods. It consists of 305,735 charged-current muon neutrino and antineutrino events detected from May 13, 2011 to May 19, 2019 (7.634~live years). These events were produced by interactions of upward-going atmospheric and astrophysical muon neutrinos and antineutrinos inside or in the medium surrounding the detector. The primary sources of background are atmospheric muons and, to a lesser extent, atmospheric electron and tau neutrinos and antineutrinos. The event selection, however, reduces non-neutrino background contamination to the $\sim 0.01\%$ level, so that the sample has a muon neutrino and antineutrino purity higher than 99.9\%.

The data are arranged in a two-dimensional histogram of reconstructed muon energy and zenith angle. The reconstructed energy range spans $E_\mu^{\rm reco} \in [500, \, 9976]~\text{GeV}$ in 13 logarithmically spaced bins of width $\Delta \log_{10}(E_\mu^{\text{reco}}) = 0.1$, while the zenith angle covers the upward-going region $\cos\theta_z \in [-1.0, \, 0.0]$ in 20 linearly spaced bins of width $\Delta\cos\theta_z = 0.05$, yielding a total of 260 bins. The angular resolution varies between $\sigma_{\cos\theta_z} \sim 0.005$ and~0.015 as a function of energy, while the energy resolution is $\sigma_{\log_{10}(E_\mu^\text{reco})} \sim 0.5$~\cite{IceCube:2020tka}.

The median opening angle of a muon track with respect to the parent muon neutrino direction is approximately $0.8^\circ \, (E_\nu/\text{TeV})^{-0.65}$ for neutrinos and $0.5^\circ \, (E_\nu/\text{TeV})^{-0.6}$ for antineutrinos between 150~GeV and 40~TeV (we obtain these by taking the median of the opening angle given by the deep-inelastic-scattering kinematics, $\theta_{\nu\mu} \approx \sqrt{2 \, M \, x \, y / \left[(1-y) \, E_\nu\right]}$, over the leading-order charged-current differential cross section~\cite{Gandhi:1995tf} with the CT18 parton distributions~\cite{Hou:2019efy}, and fitting a power law in energy). Thus, within the angular bins of this analysis, the observed zenith angle can be taken as the true zenith angle of the incident neutrino~\cite{IceCube:2020tka}. 

In contrast, there is not a one-to-one mapping between the true and reconstructed energies. While the energy estimator is biased, this does not affect the analysis. Yet, this bias implies that some reconstructed muon energies can exceed the true energy of the neutrino that originated the muon track~\cite{IceCube:2015qii}. This happens because the muon energy is estimated from the amount of light that the muon emits in the detector. At high energies, the energy losses of a muon are mainly stochastic and grow with its energy, so more energetic muons emit more light. At the lowest energies in this sample, however, muons lose energy mainly through ionization, at a rate that barely depends on their energy, so a low-energy muon emits nearly as much light as a more energetic one and can be mistaken for it~\cite{IceCube:2015qii}. Since the muon can carry most of the neutrino energy, its estimated energy can then exceed that of the neutrino.

Including the contributions from atmospheric and astrophysical neutrinos, we find that 99\% of all events were produced by neutrinos and antineutrinos with energies between approximately 150~GeV and 40~TeV. Hence, our analysis is most sensitive to this neutrino energy range. The sterile neutrino search~\cite{IceCube:2020tka} quotes a slightly different range, 130~GeV--30~TeV, based on the predicted energy distribution of conventional atmospheric neutrinos.

%%%%%%%%%%%%%%%%%%%%%%%%%%%%%%%%%%%%%%%%%%%%%%%%
\section{Neutrino fluxes}
\label{ap:nu_flux}
%%%%%%%%%%%%%%%%%%%%%%%%%%%%%%%%%%%%%%%%%%%%%%%%

Most of the events in this data sample are produced by atmospheric neutrinos, which come from cosmic-ray interactions with the atmosphere. Their flux has two components: a conventional one, arising from pion and kaon decays, whose neutrino spectrum falls as $\sim E_\nu^{-3.7}$, and a subdominant prompt one, arising from the decay of charmed hadrons, with a harder spectrum of $\sim E_\nu^{-2.7}$. The prompt component has yet to be observed experimentally, so its normalization is not known. However, it is expected to contribute at only the 0.1\% level to this sample~\cite{IceCube:2020tka}, and we therefore neglect it. In addition to other parameters that control the energy and angular dependence of this neutrino flux (Table~\ref{aptab:nuisances}), we allow the normalization with respect to the reference model, $\Phi_{\rm conv}$, and the shift in the spectral index (using $E_{\rm conv, 0} = 2.2~\text{TeV}$ as the pivot energy), $\Delta\gamma_{\rm conv}$, to vary, imposing the same Gaussian priors as in the sterile neutrino analysis~\cite{IceCube:2020phf, IceCube:2020tka}.

Another, subdominant but non-negligible, contribution comes from astrophysical neutrinos, produced by energetic sources beyond the Solar System. Depending on their spectrum, these neutrinos can become significant at the highest energies considered here. We model this flux as isotropic and following an unbroken single power law, $E_\nu^{-2.5 + \Delta\gamma_{\rm astro}}$, with the three flavors in equal proportions. The central value of the spectral index, $\gamma_{\rm astro} = 2.5$, and the reference normalization were obtained from a previous IceCube analysis~\cite{IceCube:2020acn}, and we allow the shift in the spectral index (using $E_{\rm astro, 0} = 100~\text{TeV}$ as the pivot energy), $\Delta \gamma_{\rm astro}$, and the relative normalization, $\Phi_{\rm astro}$, to vary freely under a flat prior, giving this component more freedom than the Gaussian prior adopted for the sterile neutrino search~\cite{IceCube:2020phf, IceCube:2020tka}. Our prior allows spectral indices as small as 0.5, so that the single power law can also accommodate the harder spectrum below $\sim 30~\text{TeV}$ for which recent IceCube analyses found evidence~\cite{IceCube:2025tgp, IceCube:2025ewu}.

In this work, we make use of the IceCube Monte Carlo included in the data release~\cite{DVN/9WGYQN_2024}, where the conventional component was computed using the Matrix Cascade Equation (\texttt{MCEq}) package~\cite{Fedynitch:2015zma}. This Monte Carlo contains the neutrino spectrum after propagation through the Earth (see below), used to obtain the event distribution under the assumption of the SM cross sections. Varying the neutrino and antineutrino cross sections alters the propagated spectrum in a non-trivial way. Thus, in order to perform our analyses, we first reconstruct the initial atmospheric spectrum at all energies and directions by iteratively matching the propagated spectrum to the one provided, for the given sets of parameters. In this way, we can compute the effect of varying the neutrino and antineutrino cross sections in the Monte Carlo, both due to propagation through the Earth and at detection. We have verified that the result agrees well with the spectrum obtained using \texttt{MCEq}~\cite{Fedynitch:2015zma}.

%%%%%%%%%%%%%%%%%%%%%%%%%%%%%%%%%%%%%%%%%%%%%%%%
\section{Propagation of neutrinos through the Earth}
\label{ap:Earth_prop}
%%%%%%%%%%%%%%%%%%%%%%%%%%%%%%%%%%%%%%%%%%%%%%%%

We propagate neutrinos through the Earth using the \texttt{nuSQuIDS} package~\cite{Arguelles:2021twb}, where the interior Earth density distribution is described by the Preliminary Reference Earth Model (PREM)~\cite{Dziewonski:1981xy}.

The propagation of neutrinos through the Earth is described by a set of transport equations, which account for four physical processes: the standard matter Hamiltonian, which encodes vacuum mass-mixing and coherent forward scattering off electrons via the matter potential; absorption and attenuation due to inelastic neutrino--nucleon interactions through both charged-current and neutral-current processes; energy downgrading of neutrinos following neutral-current interactions; and regeneration of tau neutrinos from tau lepton decays following charged-current interactions of tau neutrinos.

At most of the energies relevant for this analysis, neutrino flavor oscillations are suppressed, and the dominant effects are absorption and energy redistribution. The upward-going flux is attenuated as it traverses the Earth, more strongly at higher energies, because the neutrino--nucleon cross section grows with energy~\cite{Gandhi:1995tf}. Neutral-current interactions degrade the neutrino energy without absorbing it~\cite{Berezinsky:1986ij}, while charged-current interactions remove the interacting neutrino and produce a lepton of the corresponding flavor. For electron and muon neutrinos, the associated lepton is rapidly absorbed in the Earth and does not contribute to the high-energy flux. In the case of tau neutrinos, however, the produced tau lepton decays before losing a significant fraction of its energy, regenerating a lower-energy tau neutrino~\cite{Halzen:1998be} and producing secondary electron and muon neutrinos~\cite{Beacom:2001xn}. For the atmospheric flux, these effects are negligible: tau neutrinos are rarely produced in the atmosphere, few appear through oscillations at TeV energies~\cite{IceCube:2020tka}, and these effects matter only for spectra harder than the atmospheric one. Despite representing a very small correction, we include these effects for both the atmospheric and the astrophysical neutrino fluxes.

\begin{figure}[t]
\centering
    \includegraphics[width=\textwidth]{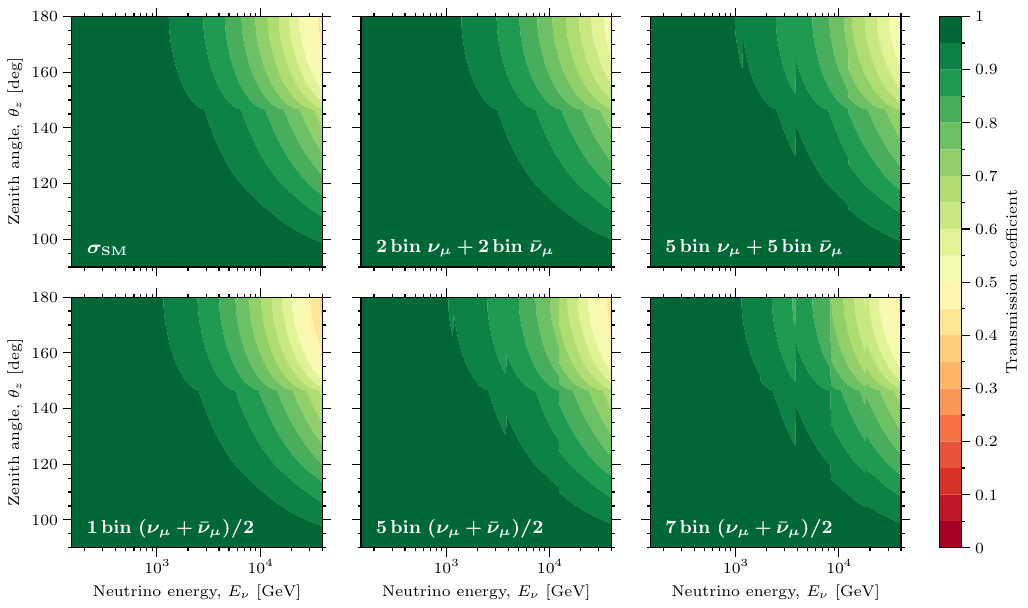}
    \caption{\textbf{Effect of Earth absorption.} Transmission coefficient (ratio of the neutrino plus antineutrino fluxes after and before crossing the Earth, not accounting for neutrino oscillations) as a function of the neutrino energy and the zenith angle of the propagation direction, for different cases. \textbf{Top-left panel:} For the SM cross sections. For the most probable values of the cross sections for five of the cases we have considered in this work: \textbf{Top-middle panel:} Two energy bins for separate neutrino and antineutrino cross sections. \textbf{Top-right panel:} Five energy bins for separate neutrino and antineutrino cross sections. \textbf{Bottom-left panel:} Single energy bin with no Gaussian prior, fixing the neutrino--antineutrino cross section ratio to the SM one. \textbf{Bottom-middle panel:} Five energy bins fixing the neutrino--antineutrino cross section ratio to the SM one. \textbf{Bottom-right panel:} Seven energy bins fixing the neutrino--antineutrino cross section ratio to the SM one. A Gaussian prior of 16\% is added to the cross-section normalizations of the first bin unless otherwise indicated.
    \label{apfig:absorption}
    }
\end{figure}

Figure~\ref{apfig:absorption} shows the combined effect of absorption and energy redistribution: the transmission coefficient (i.e., the ratio of the neutrino plus antineutrino fluxes after and before crossing the Earth, not accounting for neutrino oscillations), for the SM cross sections and for the most probable cross sections in five of our analyses (see below).

A complementary approach to the one presented in this work is absorption neutrino tomography of Earth, which uses the absorption and attenuation effects to measure the density profile of the Earth instead of the cross section. It was first performed in~\cite{Donini:2018tsg} and recently refined with additional data~\cite{IceCube:2026cgy}.

%%%%%%%%%%%%%%%%%%%%%%%%%%%%%%%%%%%%%%%%
%%%%%%%%%%%%%%%%%%%%%%%%%%%%%%%%%%%%%%%%

%%%%%%%%%%%%%%%%%%%%%%%%%%%%%%%%%%%%%%%%%%%%%%%%
\section{Parameter estimation}
\label{ap:parameter_est}
%%%%%%%%%%%%%%%%%%%%%%%%%%%%%%%%%%%%%%%%%%%%%%%%

The energy dependence of the cross sections is studied by dividing the neutrino energy range to which the data we use are sensitive, $E_\nu \in [150~\text{GeV}, \, 40~\text{TeV}]$, into several bins. For the analyses presented in the main text, the first bin, $E_1 \in [150, \, 360]~\text{GeV}$, covers the energy region where fixed-target measurements exist. The remaining bins have equal logarithmic widths. We use the CSMS cross sections~\cite{Cooper-Sarkar:2011jtt} as a reference and vary the normalization of each bin independently relative to this calculation. Since data exist for the energies in the first bin, we impose a Gaussian prior on this bin alone, while allowing the normalizations of all other bins to vary freely. Following the Super-Kamiokande prescription~\cite{Super-Kamiokande:2026xsm}, we  adopt a Gaussian prior with a 16\% standard deviation. This choice is conservative, as it roughly matches the total uncertainty in the highest-energy bin of the NuTeV analysis~\cite{NuTeV:2005wsg}, whereas uncertainties at lower energies typically amount to only a few percent~\cite{GargamelleSPS:1981hpd, Berge:1987zw, Seligman:1997fe, NuTeV:2005wsg, NOMAD:2007krq}. The impact of this prior on our results is discussed below.

%%%%%%%%%%%%%%%%%%%%%%%%%%%%%%%%%%%%%%%%
\begin{figure}
\centering
    \includegraphics[width=\textwidth]{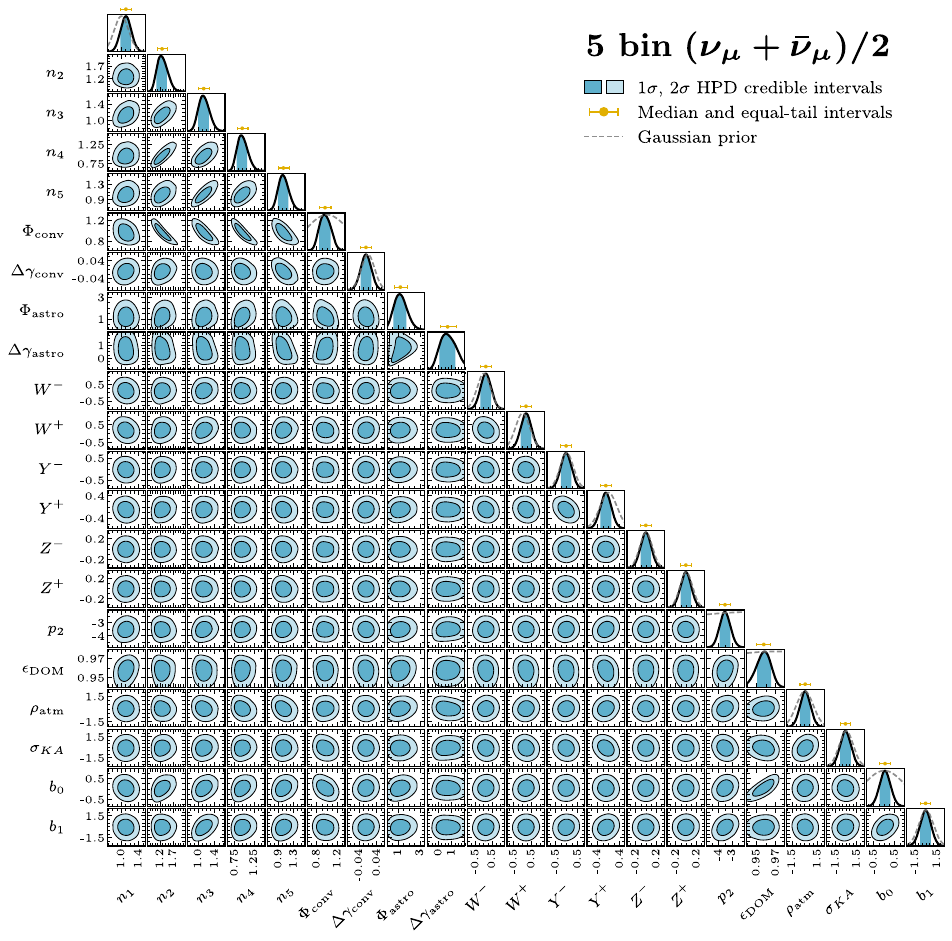}
    \caption{\textbf{Triangle plot for the average muon neutrino--antineutrino charged-current cross section analysis} corresponding to the case of five energy bins fixing the neutrino--antineutrino cross section ratio to the SM one. This corresponds to Fig.~\ref{fig:xsec_mean_result} in the main text. We show the HPD $1\sigma$ and $2\sigma$ regions for all the pairwise combinations of the 16 nuisance parameters plus 5 normalizations of the cross sections and the 21 1D posterior distributions, the latter together with their $1\sigma$ HPD and equal-tail intervals.
    \label{apfig:Tavg5}
    }
\end{figure}
%%%%%%%%%%%%%%%%%%%%%%%%%%%%%%%%%%%%%%%%

%%%%%%%%%%%%%%%%%%%%%%%%%%%%%%%%%%%%%%%%
\begin{figure}
\centering
    \includegraphics[width=\textwidth]{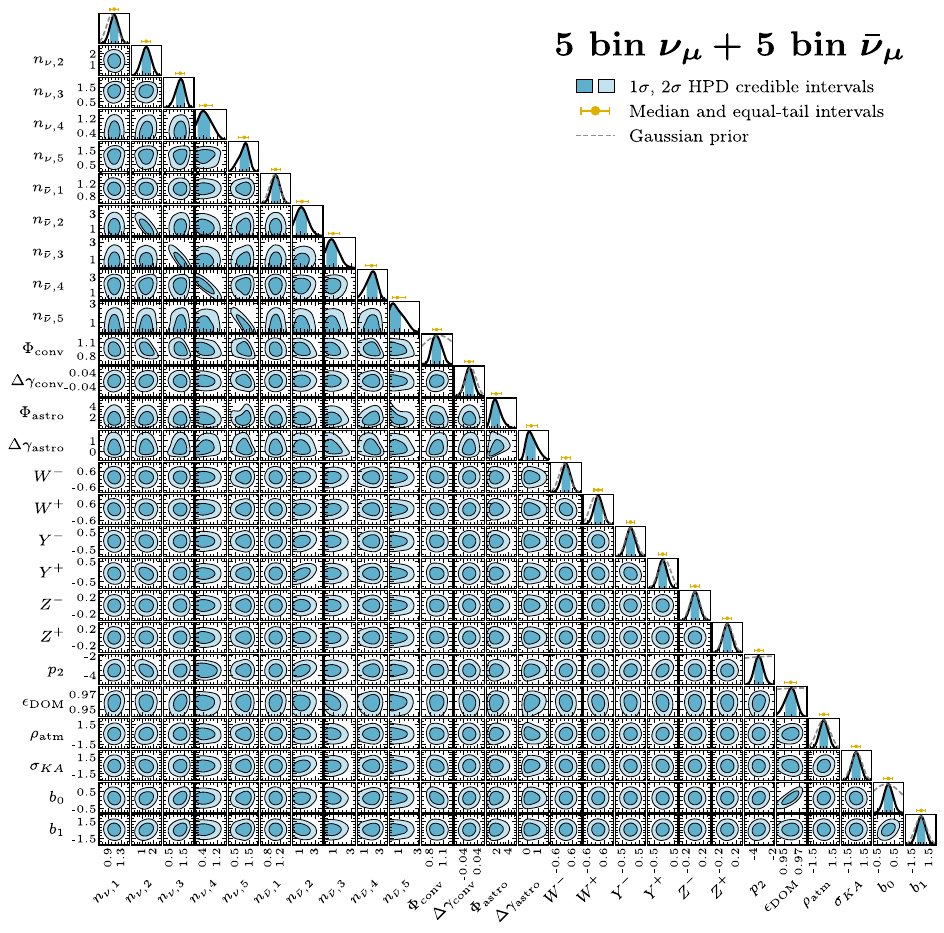}
    \caption{\textbf{Triangle plot for the separate muon neutrino and antineutrino charged-current cross sections analysis} corresponding to the case of five energy bins with freely varying normalizations of the neutrino and antineutrino cross sections. This corresponds to Fig.~\ref{fig:xsec_nunb_result} in the main text. Regions and intervals are shown as in Fig.~\ref{apfig:Tavg5}.
    \label{apfig:Tnnb5}
    }
\end{figure}
%%%%%%%%%%%%%%%%%%%%%%%%%%%%%%%%%%%%%%%%

All the events in the data sample are characterized by their reconstructed muon energy and cosine of their zenith angle. We use all 260 bins in the dataset and perform a Bayesian analysis based on a likelihood defined as the product over bins of the Poisson probability of observing $N^{\rm data}_i$ events in bin $i$ given the predicted expectation $N^{\rm th}_i$, multiplied by the priors on the nuisance parameters, $\boldsymbol{\eta}$ (see below). The corresponding log-likelihood (up to a constant) is defined as
\begin{equation}
\label{eq:L}
\ln \mathcal{L} (\boldsymbol{n_\sigma} \, ; \boldsymbol{\eta}) = \sum_{i\in {\rm bins}}  \left(N^{\rm data}_i \, \ln N^{\rm th}_i ({\boldsymbol n_\sigma} \, ; \boldsymbol{\eta}) - N^{\rm th}_i (\boldsymbol{n_\sigma} \, ; {\boldsymbol \eta})\right) - \sum_j \frac{(\eta_j-\eta^0_j)^2}{2 \, \sigma_j^2} ~.
\end{equation}
The expected number of events in bin $i$ for a given set of cross-section normalizations in different bins, $\boldsymbol{n_\sigma}$, is given by $N^{\rm th}_i ({\boldsymbol n_\sigma} \, ; {\boldsymbol \eta})$ and $N^{\rm data}_i$ is the number of detected events in bin $i$. The index $j$ runs over the nuisance parameters with Gaussian priors, of mean $\eta^0_j$ and standard deviation $\sigma_j$; the two astrophysical-flux parameters have flat priors (Table~\ref{aptab:nuisances}). For each set of parameter values, the neutrino and antineutrino fluxes are propagated from the atmosphere to the detector, and the resulting flux --- a function of the true neutrino energy $E_\nu$ and zenith angle $\theta_z$ --- is used to reweight the IceCube Monte Carlo events. The reweighted events are then binned in two dimensions as a function of the reconstructed muon energy $E_\mu^{\text{reco}}$ and cosine of zenith angle $\cos \theta_z^\text{reco}$, using 13 bins in reconstructed muon energy and 20 in cosine of zenith angle. All the nuisance parameters are continuously varied and marginalized simultaneously with the normalizations of the cross sections with respect to the CSMS theoretical estimate~\cite{Cooper-Sarkar:2011jtt} in the selected bins for each analysis.

All the credible intervals we depict in this work correspond to 68.27\% highest probability density (HPD) intervals around the mode, which we also show as a reference. They are obtained from one-dimensional marginalized posteriors, which include the effect of the correlations with all other parameters but do not display them. The joint posterior distributions and parameter correlations for the analyses described in the main text are shown in the triangle plots in Figs.~\ref{apfig:Tavg5} and~\ref{apfig:Tnnb5} and the mode of the posteriors and the 68.27\% HPD intervals for all parameters are also provided in tabular format in Tables~\ref{aptab:nuisances} and~\ref{aptab:xsec}.

%%%%%%%%%%%%%%%%%%%%%%%%%%%%%%%%%%%%%%%%
\begin{table}
	\centering
    \resizebox{\textwidth}{!}{
	\begin{tabular}{@{\extracolsep{\fill}}lccccccc}
		\hline
		Parameter & Prior & Average & \textbf{Average} & Average & Average & $\nu/\bar\nu$ & $\mathbf{\nu/\bar\nu}$ \\
 		& & 1 bin, no prior & \textbf{5 bins} & 7 bins & 5 bins, no prior & 2 bins & \textbf{5 bins} \\[2ex]
        \hline 
        \multicolumn{8}{c}{\textbf{Detector parameters}} \\
		\hline \\
        DOM efficiency ($\epsilon_{\rm DOM}$) & $\mathcal{N}(0.97,0.10)$ & $0.961^{+0.006}_{-0.007}$ & $\mathbf{0.958^{+0.007}_{-0.008}}$ & $0.959^{+0.007}_{-0.008}$ & $0.966 \pm 0.008$ & $0.964 \pm 0.007$ & $\mathbf{0.960^{+0.007}_{-0.008}}$ \\[1ex]
        Bulk ice gradient 0 ($b_{0}$) & $\mathcal{N}(0.0,1.0)$ & $-0.08^{+0.24}_{-0.23}$ & $\mathbf{0.03^{+0.27}_{-0.26}}$ & $0.04 \pm 0.26$ & $0.06 \pm 0.27$ & $-0.10^{+0.26}_{-0.25}$ & $\mathbf{0.13^{+0.26}_{-0.27}}$ \\[1ex]
		Bulk ice gradient 1 ($b_{1}$) & $\mathcal{N}(0.0,1.0)$ & $0.3 \pm 0.5$ & $\mathbf{-0.1^{+0.7}_{-0.8}}$ & $0.0^{+0.7}_{-0.8}$ & $-0.1 \pm 0.7$ & $0.4 \pm 0.7$ & $\mathbf{0.1 \pm 0.8}$ \\[1ex]
        Forward ice hole ($p_2$) & $\mathcal{N}(-1.0,10.0)$ & $-3.3 \pm 0.4$ & $\mathbf{-3.5 \pm 0.4}$ & $-3.5 \pm 0.4$ & $-3.4 \pm 0.4$ & $-3.5 \pm 0.4$ & $\mathbf{-3.5^{+0.5}_{-0.4}}$ \\[2ex]
        \hline 
        \multicolumn{8}{c}{\textbf{Conventional flux parameters}} \\
        \hline \\
		Normalization ($\Phi_{\rm conv}$) & $\mathcal{N}(1.0,0.4)$ & $0.92^{+0.13}_{-0.11}$ & $\mathbf{0.95^{+0.12}_{-0.10}}$ & $0.93^{+0.13}_{-0.08}$ & $0.72^{+0.13}_{-0.10}$ & $1.11^{+0.13}_{-0.10}$ & $\mathbf{0.93^{+0.12}_{-0.09}}$ \\[1ex]
		Spectral shift ($\Delta\gamma_{\rm conv}$) & $\mathcal{N}(0.00,0.03)$ & $-0.067^{+0.012}_{-0.011}$ & $\mathbf{-0.009^{+0.025}_{-0.024}}$ & $-0.008^{+0.024}_{-0.025}$ & $0.005^{+0.024}_{-0.025}$ & $-0.055^{+0.014}_{-0.015}$ & $\mathbf{-0.009^{+0.024}_{-0.025}}$ \\[1ex]
        Atmospheric density ($\rho_{\rm atm}$) & $\mathcal{N}(0.0,1.0)$ & $-0.5 \pm 0.6$ & $\mathbf{0.0 \pm 0.6}$ & $0.0 \pm 0.6$ & $0.1^{+0.7}_{-0.6}$ & $-0.2 \pm 0.6$ & $\mathbf{0.1 \pm 0.7}$ \\[1ex]
        Barr $W^-$ & $\mathcal{N}(0.0,0.4)$ & $-0.03^{+0.32}_{-0.31}$ & $\mathbf{0.12^{+0.33}_{-0.32}}$ & $0.12^{+0.32}_{-0.33}$ & $0.02 \pm 0.33$ & $-0.10 \pm 0.33$ & $\mathbf{0.12^{+0.34}_{-0.33}}$ \\[1ex]
		Barr $W^+$ & $\mathcal{N}(0.0,0.4)$ & $-0.08^{+0.27}_{-0.28}$ & $\mathbf{0.20 \pm 0.30}$ & $0.19 \pm 0.31$ & $0.02 \pm 0.31$ & $-0.18 \pm 0.29$ & $\mathbf{0.18 \pm 0.31}$ \\[1ex]
		Barr $Y^-$ & $\mathcal{N}(0.0,0.3)$ & $-0.06^{+0.25}_{-0.23}$ & $\mathbf{-0.03^{+0.25}_{-0.24}}$ & $-0.02^{+0.24}_{-0.25}$ & $-0.06 \pm 0.24$ & $-0.09^{+0.25}_{-0.24}$ & $\mathbf{-0.02 \pm 0.25}$ \\[1ex]
		Barr $Y^+$ & $\mathcal{N}(0.0,0.3)$ & $-0.14 \pm 0.19$ & $\mathbf{-0.10^{+0.20}_{-0.19}}$ & $-0.09^{+0.20}_{-0.19}$ & $-0.16^{+0.19}_{-0.20}$ & $-0.22^{+0.20}_{-0.19}$ & $\mathbf{-0.11^{+0.22}_{-0.21}}$ \\[1ex]
		Barr $Z^-$ & $\mathcal{N}(0.00,0.12)$ & $-0.01 \pm 0.10$ & $\mathbf{0.00 \pm 0.10}$ & $0.00^{+0.11}_{-0.10}$ & $0.00 \pm 0.10$ & $-0.01^{+0.10}_{-0.11}$ & $\mathbf{0.00^{+0.11}_{-0.10}}$ \\[1ex]
		Barr $Z^+$ & $\mathcal{N}(0.00,0.12)$ & $-0.03 \pm 0.10$ & $\mathbf{-0.01 \pm 0.10}$ & $-0.01 \pm 0.10$ & $0.00 \pm 0.10$ & $-0.02^{+0.10}_{-0.11}$ & $\mathbf{-0.01^{+0.11}_{-0.10}}$ \\[1ex]
        Kaon re-interaction $(\sigma_{KA})$ & $\mathcal{N}(0.0,1.0)$ & $-0.1 \pm 0.8$ & $\mathbf{-0.1 \pm 0.8}$ & $-0.1 \pm 0.8$ & $-0.1 \pm 0.8$ & $-0.3 \pm 0.8$ & $\mathbf{-0.1 \pm 0.8}$ \\[2ex]
        \hline
        \multicolumn{8}{c}{\textbf{Astrophysical flux parameters}} \\
        \hline \\
		Normalization ($\Phi_{\rm astro}$) & $\mathcal{U}(0,10)$ & $1.5 \pm 0.5$ & $\mathbf{1.1 \pm 0.6}$ & $1.2^{+0.7}_{-0.6}$ & $1.1^{+0.6}_{-0.4}$ & $1.7^{+0.9}_{-0.7}$ & $\mathbf{1.6^{+0.9}_{-0.7}}$ \\[1ex]
		Spectral shift ($\Delta\gamma_{\rm astro}$) & $\mathcal{U}(-2,2)$ & $0.9^{+0.6}_{-0.5}$ & $\mathbf{0.6^{+0.7}_{-0.5}}$ & $0.6^{+0.8}_{-0.5}$ & $0.0^{+0.5}_{-0.3}$ & $1.1 \pm 0.5$ & $\mathbf{0.2^{+0.6}_{-0.4}}$ \\[1ex]
		\hline
	\end{tabular}%
    }
	\caption{\textbf{Nuisance parameters for the analyses in this work.} Mode of the posterior and the 68.27\% HPD interval for all the nuisance parameters included in all the analyses in this work, classified as detector, conventional flux or astrophysical flux parameters. We also indicate the priors for all the nuisance parameters used in all analyses. The two cases discussed in the main text are in bold font. The distributions $\mathcal{N}$ (normal) and $\mathcal{U}$ (uniform) are defined as in Table~\ref{tab:normalization_1_5} in the main text. For a detailed description of these parameters, we refer the reader to the original IceCube analyses~\cite{IceCube:2020phf, IceCube:2020tka}.}
	\label{aptab:nuisances}
\end{table}
%%%%%%%%%%%%%%%%%%%%%%%%%%%%%%%%%%%%%%%%

%%%%%%%%%%%%%%%%%%%%%%%%%%%%%%%%%%%%%%%%
\begin{table}
	\centering
	\begin{tabular*}{\columnwidth}{@{\extracolsep{\fill}}lcccccc}
	\hline
	Parameter & Average & \textbf{Average} & Average & Average & $\nu/\bar\nu$ & $\mathbf{\nu/\bar\nu}$ \\
 	& 1 bin, no prior & \textbf{5 bins} & 7 bins & 5 bins, no prior & 2 bins & \textbf{5 bins} \\[1ex]
	\hline \\
	$n_{1}$ & $1.12^{+0.16}_{-0.13}$ & $\mathbf{1.11^{+0.12}_{-0.13}}$ & $1.11 \pm 0.13$ & $1.7^{+0.5}_{-0.4}$ & -- & -- \\[1ex]
	$n_{2}$ & -- & $\mathbf{1.22^{+0.24}_{-0.17}}$ & $1.27^{+0.31}_{-0.27}$ & $1.45^{+0.29}_{-0.23}$ & -- & -- \\[1ex]
	$n_{3}$ & -- & $\mathbf{1.11^{+0.15}_{-0.12}}$ & $1.14^{+0.34}_{-0.31}$ & $1.45^{+0.26}_{-0.21}$ & -- & -- \\[1ex]
	$n_{4}$ & -- & $\mathbf{0.92^{+0.15}_{-0.12}}$ & $1.22^{+0.35}_{-0.39}$ & $1.12^{+0.20}_{-0.17}$ & -- & -- \\[1ex]
	$n_{5}$ & -- & $\mathbf{1.01^{+0.15}_{-0.13}}$ & $0.85^{+0.33}_{-0.31}$ & $1.25^{+0.22}_{-0.19}$ & -- & -- \\[1ex]
	$n_{6}$ & -- & -- & $1.05^{+0.25}_{-0.26}$ & -- & -- & -- \\[1ex]
	$n_{7}$ & -- & -- & $0.98^{+0.19}_{-0.18}$ & -- & -- & -- \\[1ex]
	\hline \\
	$n_{\nu,1}$ & -- & -- & -- & -- & $1.07^{+0.13}_{-0.12}$ & $\mathbf{1.08 \pm 0.13}$ \\[1ex]
	$n_{\nu,2}$ & -- & -- & -- & -- & $1.17 \pm 0.15$ & $\mathbf{1.4 \pm 0.4}$ \\[1ex]
	$n_{\nu,3}$ & -- & -- & -- & -- & -- & $\mathbf{1.27^{+0.28}_{-0.33}}$ \\[1ex]
	$n_{\nu,4}$ & -- & -- & -- & -- & -- & $\mathbf{0.40^{+0.39}_{-0.30}}$ \\[1ex]
	$n_{\nu,5}$ & -- & -- & -- & -- & -- & $\mathbf{1.21^{+0.28}_{-0.41}}$ \\[1ex]
	\hline \\
	$n_{\bar\nu,1}$ & -- & -- & -- & -- & $1.04 \pm 0.14$ & $\mathbf{1.03 \pm 0.14}$ \\[1ex]
	$n_{\bar\nu,2}$ & -- & -- & -- & -- & $0.31^{+0.37}_{-0.26}$ & $\mathbf{1.2^{+0.8}_{-0.7}}$ \\[1ex]
	$n_{\bar\nu,3}$ & -- & -- & -- & -- & -- & $\mathbf{0.7^{+0.7}_{-0.6}}$ \\[1ex]
	$n_{\bar\nu,4}$ & -- & -- & -- & -- & -- & $\mathbf{2.1^{+0.6}_{-0.8}}$ \\[1ex]
	$n_{\bar\nu,5}$ & -- & -- & -- & -- & -- & $\mathbf{0.4^{+0.8}_{-0.4 \,\, \dagger}}$ \\[1ex]
	\hline
	\end{tabular*}
	\caption{\textbf{Cross-section normalizations for the analyses in this work.}	Mode of the posterior and the 68.27\% HPD interval for the normalizations of the cross sections for the analyses in this work. For all the analyses, except for those where it is explicitly stated ``no prior", we impose a Gaussian prior of 16\% on the cross-section normalization of the first bin, $E_1 \in [150, \, 360]~\text{GeV}$. The two cases discussed in the main text are in bold font. Explicitly, $n_k$: average $\nu+\bar\nu$ analyses fixing the neutrino--antineutrino cross section ratio to the SM one; $n_{\nu,k}$, $n_{\bar\nu,k}$: separate $\nu$ and $\bar\nu$ analyses. Bin $k$ covers a different energy range in each case. For the single-bin analysis: $E_1 \in [150~\text{GeV}, \, 40~\text{TeV}]$. For two bins: $E_1 \in [150, \, 360]~\text{GeV}$ and $E_2 \in [0.36, \, 40]~\text{TeV}$. For five bins: $E_1 \in [150, \, 360]~\text{GeV}$, $E_2 \in [0.36, \, 1.17]~\text{TeV}$, $E_3 \in [1.17, \, 3.79]~\text{TeV}$, $E_4 \in [3.79, \, 12.32]~\text{TeV}$, and $E_5 \in [12.32, \, 40]~\text{TeV}$. For seven bins: $E_1 \in [150, \, 360]~\text{GeV}$, $E_2 \in [360, \, 789]$~GeV, $E_3 \in [0.789, \, 1.73]~\text{TeV}$, $E_4 \in [1.73, \, 3.79]~\text{TeV}$, $E_5 \in [3.79, \, 8.32]$~\text{TeV}, $E_6 \in [8.32, \, 18.24]~\text{TeV}$, and $E_7 \in [18.24, \, 40]~\text{TeV}$. \\ $^\dagger$ The interval reaches the physical bound at 0.}
    \label{aptab:xsec}
\end{table}
%%%%%%%%%%%%%%%%%%%%%%%%%%%%%%%%%%%%%%%%

%%%%%%%%%%%%%%%%%%%%%%%%%%%%%%%%%%%%%%%%%%%%%%%%
\section{Systematic uncertainties}
\label{ap:systematics}
%%%%%%%%%%%%%%%%%%%%%%%%%%%%%%%%%%%%%%%%%%%%%%%%

In addition to the signal simulation, the IceCube Monte Carlo also includes the treatment of systematic uncertainties~\cite{DVN/9WGYQN_2024}. Their impact is modeled via nuisance parameters, each encoding a continuous parameterization of a different effect that modifies the Monte Carlo event weights accordingly. These nuisance parameters are presented in two ways: gradients and splines, but not every parameter is available as a spline. For a detailed description of the treatment of systematic uncertainties, we refer the reader to the original analyses~\cite{IceCube:2020phf, IceCube:2020tka}.

The dominant sources of systematic uncertainty in these analyses are the normalization and spectral shape of the atmospheric and astrophysical neutrino fluxes, the bulk optical properties of the South Pole ice, and the local response of the IceCube digital optical modules (DOMs). Additional subdominant uncertainties associated with the primary cosmic-ray flux and the hadronic interaction models governing air-shower development are also included. In total, we include 16 nuisance parameters. Other potential sources of uncertainty, such as the contributions from the prompt atmospheric neutrino flux, from atmospheric muons, from misidentified conventional atmospheric electron neutrinos, uncertainties on the Earth density profile, and a potential neutrino-to-antineutrino ratio of the astrophysical flux different from one, were investigated and found to be subdominant compared to statistical and other systematic uncertainties of the dataset~\cite{IceCube:2020phf, IceCube:2020tka}, and have a negligible impact on the results presented in this work. They have been neglected.

Given the nuisance parameter priors listed in Table~\ref{aptab:nuisances}, systematic uncertainties represent the primary source of error in the current dataset. To evaluate the relative effect of each systematic uncertainty on the normalization of the cross sections in each energy bin, we have analyzed their impact directly from the joint Markov chain Monte Carlo (MCMC) posterior samples we generated to obtain our results. For a given systematic source, its impact on the cross-section normalization in each energy bin is defined as the change in the average value of that normalization when the nuisance parameter is held near the limits of its 68.27\% credible interval, while fully accounting for variations in all other parameters. The normalization and spectral shape of the atmospheric neutrino flux are found to be the dominant sources of systematic uncertainty on the derived cross-section normalizations. Next in importance are the normalization and shape of the astrophysical neutrino flux and the properties of the bulk ice in the array. Altogether, these contributions account for more than 90\% of the overall systematic uncertainty in general. When fitting the neutrino and antineutrino cross sections separately, the photon detection efficiency of the DOMs also contributes significantly to the systematic error budget in the highest energy bin.

%%%%%%%%%%%%%%%%%%%%%%%%%%%%%%%%%%%%%%%%%%%%%%%%
\section{Robustness checks}
\label{ap:robustness}
%%%%%%%%%%%%%%%%%%%%%%%%%%%%%%%%%%%%%%%%%%%%%%%%

%%%%%%%%%%%%%%%%%%%%%%%%%%%%%%%%%%%%%%%%
\begin{figure}
\centering
    \includegraphics[width=\textwidth]{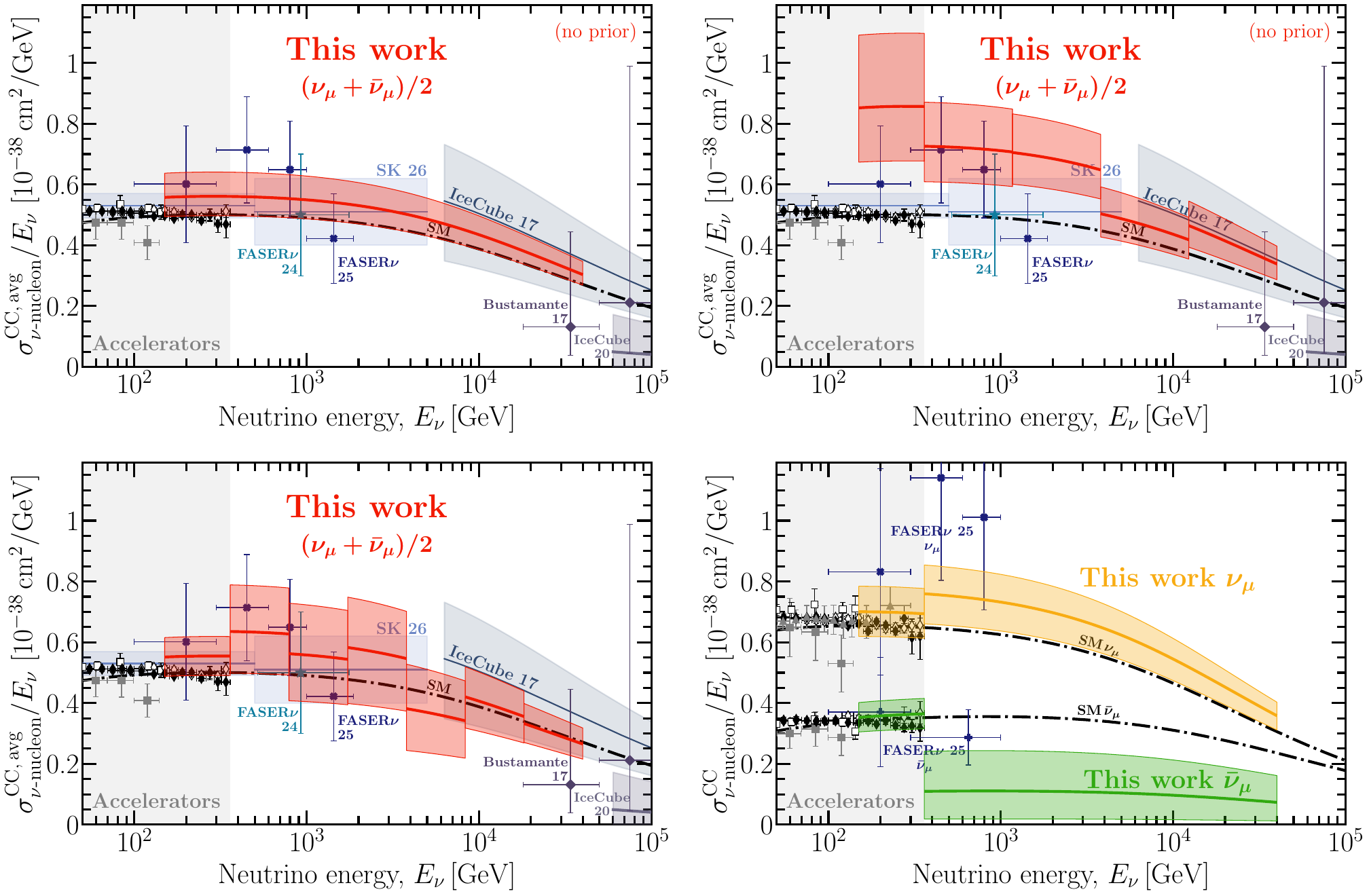}
    \caption{\textbf{Measurement of the muon neutrino charged-current cross section}. \textbf{Top-left panel:} Average neutrino--antineutrino cross section, as in Fig.~\ref{fig:xsec_mean_result}, but for a single energy bin and without using a Gaussian prior on the cross-section normalization. \textbf{Top-right panel:} As in Fig.~\ref{fig:xsec_mean_result}, but without a Gaussian prior on the cross-section normalization for the first bin. \textbf{Bottom-left panel:} As in Fig.~\ref{fig:xsec_mean_result}, but for seven energy bins. \textbf{Bottom-right panel:} Muon neutrino and antineutrino cross sections, as in Fig.~\ref{fig:xsec_nunb_result}, but for two energy bins.
    \label{apfig:xsec2x2}
    }
\end{figure}
%%%%%%%%%%%%%%%%%%%%%%%%%%%%%%%%%%%%%%%%

To assess the robustness of our results and further clarify them, we have performed several checks with modified analysis setups.

We first consider a single bin over the entire energy range ($E_\nu \in [150~\text{GeV}, \, 40~\text{TeV}]$) and determine the overall normalization of the average muon neutrino--antineutrino cross section with respect to the CSMS calculation~\cite{Cooper-Sarkar:2011jtt}. Unlike our main analyses, which constrain the first bin, from 150 to 360~GeV, with a Gaussian prior based on fixed-target measurements, this analysis imposes no Gaussian prior on the cross section, so that no information from other cross section measurements enters the fit. The top-left panel of Fig.~\ref{apfig:xsec2x2} depicts this result. The central value is similar to that of the first bin in the five-bin analysis including a Gaussian prior, Fig.~\ref{fig:xsec_mean_result}, and the uncertainty is just slightly larger. This is the result of the tension between the low-energy and high-energy data. The former prefer a higher normalization of the cross section, whereas the latter a lower one.

The tension becomes clear when the five-bin analysis of Fig.~\ref{fig:xsec_mean_result} is repeated without the Gaussian prior on the first bin (top-right panel of Fig.~\ref{apfig:xsec2x2}). The data then prefer a higher cross section in that bin, with a larger uncertainty, approximately $2\sigma$ above the theoretical estimate. This implies a lower normalization of the conventional atmospheric neutrino flux, which raises the cross-section normalization in all energy bins, with slightly larger uncertainties. This is also favored by the constraining prior on the spectral index of the atmospheric flux. At high energies, however, absorption, which adds a further dependence on the cross section, prefers a lower normalization. Overall, the average cross section across two decades in energy is larger than with the prior. Therefore, the prior on the first bin helps constrain the normalization of the atmospheric neutrino flux and, through it, the cross section in all energy bins.

Furthermore, to compare with the results in Fig.~\ref{fig:xsec_mean_result} of the main text, we have also performed an analysis with seven energy bins, including the same Gaussian prior for the first bin, chosen to have the same width in both analyses. The results are presented in the bottom-left panel of Fig.~\ref{apfig:xsec2x2}. They are similar to the case with five bins, albeit with larger uncertainties. We have estimated whether adding two extra bins is required to better describe the data by using the Watanabe-Akaike Information Criterion (WAIC)~\cite{Watanabe2010}. With a $\Delta$WAIC-to-standard-error ratio of 3.3, the analysis reveals that, with current data, the finer binning in the seven-bin model fails to add any further descriptive power.   

Next, we consider the separate measurement of the neutrino and antineutrino cross sections, but with a simpler model. We define two energy bins for each cross section, imposing a Gaussian prior on their first bin ($E_1 \in [150, \, 360]~\text{GeV}$), as done for other cases. The results are shown in the bottom-right panel of Fig.~\ref{apfig:xsec2x2}. Comparing the marginalized posteriors of this analysis with those obtained by setting the antineutrino cross-section normalization in the second bin to zero shows no preference for a non-zero antineutrino cross section, with a $p$-value of 0.22. This corresponds to $0.8\sigma$ (one-sided) significance, assuming the log-likelihood ratio behaves as an equal mixture of a delta function at zero and a
$\chi^2$ distribution for one degree of freedom~\cite{Chernoff1954}. Therefore, the five-bin case, discussed in the main text, 
%in which we split both cross sections in five energy bins, 
provides a stronger hint of a measurement of a non-zero antineutrino cross section. Conversely, driven by the low antineutrino cross section result, the neutrino–antineutrino separation reaches $3.3\sigma$ in the two-bin case when evaluated with the Mahalanobis distance of the highest-energy bin.

The mode of the posteriors and the 68.27\% HPD intervals for all parameters in all these analyses are also provided in tabular format in Tables~\ref{aptab:nuisances} and~\ref{aptab:xsec}.

Finally, consistent with the IceCube Monte Carlo approach, we assume isoscalar nuclear targets throughout this work — containing equal numbers of protons and neutrons — for both neutrino propagation through the Earth and neutrino detection. In reality, however, nuclei are not strictly isoscalar. This approximation has the largest impact at lower energies, where the valence-quark contribution dominates and neutrino and antineutrino cross sections on protons and neutrons differ appreciably. Therefore, we evaluated the impact of this assumption on our results by performing dedicated runs accounting for the different number of protons and neutrons in nuclear targets, both at propagation and at detection. We found the effect to be at the level of a few percent at most, shifting the mode and credible intervals by less than $0.1\sigma$. Although small at present, accounting for this correction may become relevant with future datasets, provided that systematic uncertainties are also reduced.

%%%%%%%%%%%%%%%%%%%%%%%%%%%%%%%%%%%%%%%%%%%%%%%%
\section{Forecast for ten times higher statistics}
\label{ap:forecast}
%%%%%%%%%%%%%%%%%%%%%%%%%%%%%%%%%%%%%%%%%%%%%%%%

\begin{figure}
\centering
    \includegraphics[width=\textwidth]{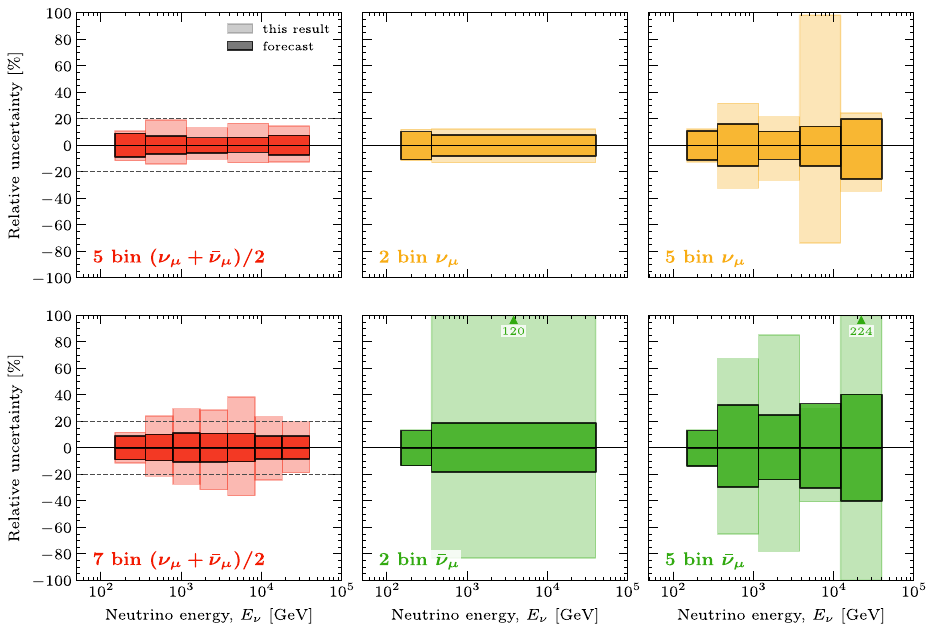}
    \caption{\textbf{Forecast for ten times higher statistics than the current dataset.} Relative uncertainties of the normalization of the neutrino and antineutrino cross sections with the current dataset (light colors) and with ten times higher statistics assuming the SM cross sections (dark colors). The results are shown for different analyses. \textbf{Top-left panel:} Five energy bins fixing the neutrino--antineutrino cross section ratio to the SM one. \textbf{Top-middle panel:} Normalizations of the neutrino cross section for two energy bins for separate neutrino and antineutrino cross sections. \textbf{Top-right panel:} As in the top-middle panel, but for five energy bins. \textbf{Bottom-left panel:} Seven energy bins fixing the neutrino--antineutrino cross section ratio to the SM one. \textbf{Bottom-middle panel:} Normalizations of the antineutrino cross section for two energy bins for separate neutrino and antineutrino cross sections. \textbf{Bottom-right panel:} As in the bottom-middle panel, but for five energy bins. A Gaussian prior of 16\% is added to the cross-section normalizations of the first bin in all cases. In those bins for which the uncertainty is larger than 100\%, the error is explicitly indicated.
    \label{apfig:forecast}
    }
\end{figure}

We project future sensitivity by performing forecast analyses on a mock dataset ten times larger than the current sample. To generate these data, we assume the true neutrino cross sections to be the SM ones, and to perform the analysis, we use identical systematic uncertainties. The results are shown in Fig.~\ref{apfig:forecast}, where we compare the relative uncertainties obtained with current data to the projected ones. We depict results for the five-bin and seven-bin analyses of the average cross section (fixing the neutrino--antineutrino cross section ratio to the SM one) and for the two-bin and five-bin analyses fitting neutrino and antineutrino cross sections separately. In all cases, we include a Gaussian prior of 16\% for the first bin from previous fixed-target measurements~\cite{GargamelleSPS:1981hpd, Berge:1987zw, Seligman:1997fe, NuTeV:2005wsg, NOMAD:2007krq}. With higher statistics, the uncertainties would decrease significantly across all cases, since the larger dataset would further constrain the nuisance parameters. Yet, the overall uncertainties would remain primarily dominated by the strong anti-correlation between the atmospheric flux normalization and those of the neutrino and antineutrino cross sections. Consequently, tightening the prior on the overall normalization of the atmospheric neutrino flux could significantly improve the precision with which neutrino and antineutrino cross sections are determined at TeV energies.

Remarkably, while current data yield only a mild $1.6\sigma$ indication of a non-zero antineutrino cross section for $E_\nu > 360~\text{GeV}$ (considering five bins per species), a tenfold increase in statistics could establish robust evidence for a non-zero antineutrino cross section at over $7\sigma$, even with the very conservative prior used for the normalization of the atmospheric flux. As mentioned above, this is partly because a larger dataset could tighten the bounds on the nuisance parameters, thereby mitigating systematic uncertainties. As indicated in the main text, these estimates of the significance assume the log-likelihood ratio to behave as a weighted mixture of $\chi^2$ distributions with up to four degrees of freedom~\cite{Chernoff1954}.

%%%%%%%%%%%%%%%%%%%%%%%%%%%%%%%%%%%%%%%%
%%%%%%%%%%%%%%%%%%%%%%%%%%%%%%%%%%%%%%%%

%%%%%%%%%%%%%%%%%%%%%%%%%%%%%%%%%%%%%%%%%%%%%%%%%%%%%%%%%%%%%%%%%%%%%%%%%%%%%%
\clearpage
\twocolumngrid

\bibliographystyle{JHEP}
\bibliography{biblio}

\end{document}